\documentclass[lettersize,journal]{IEEEtran}
\usepackage{amsmath,amsfonts, amssymb}
\usepackage{array}
\usepackage[caption=false,font=normalsize,labelfont=sf,textfont=sf]{subfig}
\usepackage{textcomp}
\usepackage{stfloats}
\usepackage{url}
\usepackage{verbatim}
\usepackage{graphicx}
\usepackage{cite}
\usepackage{physics}
\usepackage{svg}
\usepackage{booktabs}
\usepackage{xcolor}
\usepackage[hidelinks]{hyperref}
\usepackage{mathtools}
\usepackage{siunitx}
\usepackage[english]{babel}
\usepackage{algpseudocodex}
\usepackage{enumitem}
\usepackage[switch]{lineno}

\DeclarePairedDelimiter\ceil{\lceil}{\rceil}

\def\multiset#1#2{\ensuremath{\left(\kern-.3em\left(\genfrac{}{}{0pt}{}{#1}{#2}\right)\kern-.3em\right)}}

\begin{document}
\title{Adaptive Policies for Resource Generation in a Quantum Network}

\author{Aksel Tacettin*, 
        Tianchen Qu*,
        Bethany Davies*, \\
        Boris Goranov, 
        Ioana-Lisandra Draganescu,
        and Gayane Vardoyan
\thanks{
*These authors contributed equally.\\
Corresponding author: Gayane Vardoyan (gvardoyan@cs.umass.edu).}%
\thanks{All authors are with EEMCS, Delft University of Technology, Delft, the Netherlands. Bethany Davies is with QuTech and Kavli Institute of Nanoscience, Delft University of Technology. Gayane Vardoyan is with MCICS, University of Massachusetts, Amherst, USA.}%
}




\maketitle

\begin{abstract}
Protocols for distributed quantum systems commonly require the simultaneous availability of $n$ entangled states, each with a fidelity above some fixed minimum $F_{\mathrm{app}}$ relative to the target maximally-entangled state. 
However, the fidelity of entangled states degrades over time while in memory. 
Entangled states are therefore rendered useless when their fidelity falls below $F_{\mathrm{app}}$. 
This is problematic when entanglement generation is probabilistic and attempted in a sequential manner, because the expected completion time until $n$ entangled states are available can be large. 
Motivated by existing entanglement generation schemes, we consider a system where the entanglement generation parameters (the success probability $p$ and fidelity $F$ of the generated entangled state) may be adjusted at each time step. 
We model the system as a Markov decision process, where the policy dictates which generation parameters $(p,F)$ to use for each attempt. 
We use dynamic programming to derive optimal policies that minimise the expected time until $n$ entangled states are available with fidelity greater than $F_{\mathrm{app}}$. 
We observe that the advantage of our optimal policies over the selected baselines increases significantly with $n$. 
In both parameter regimes explored, which are based closely on current experiments, we find that the optimal policy can provide a speed-up of as much as a factor of twenty over a constant-action policy.
In addition, we propose a simple heuristic method to compute policies that perform either optimally or near-optimally in the parameter regimes explored. We verify empirically that our heuristic method can be used to find high-performing policies in parameter regimes where finding an optimal policy is intractable.
\end{abstract}


\section{Introduction}
\begin{figure}[t]
    \centering
    \includegraphics[width=0.8\linewidth]{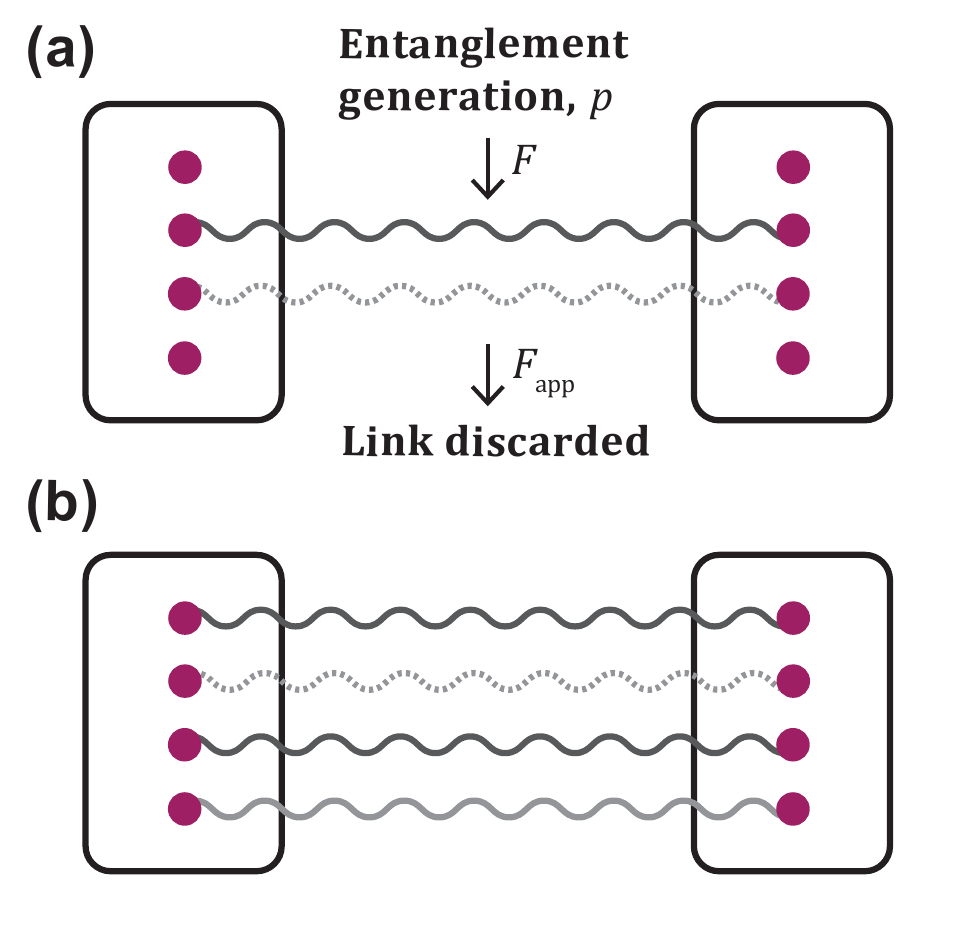}
    \caption{ \textbf{(a) Depiction of an intermediate state with $n=4$.} At the beginning of each time step, the system chooses generation parameters $(p,F)\in \mathcal{A}$. Then, an entanglement generation attempt is carried out, which with probability $p$ generates a link with fidelity $F$. If successful, the generated link is immediately transferred to memory. While in memory, the link is subject to decoherence. If the link's fidelity falls below $F_{\mathrm{app}}$, it is discarded. We are interested in the case where the system starts with no links in memory. It will then pass through a number of intermediate states, until the first time $T$ when it reaches a terminal state that has $n$ links simultaneously available in memory. 
    \textbf{(b) Depiction of a terminal state with $n=4$.} 
    }
\label{fig:entanglement}
\end{figure}

\IEEEPARstart{P}{rotocols} for distributed quantum systems commonly require multiple entangled pairs of qubits, also referred to as \textit{entangled links}, or just \textit{links}.   
Examples of protocols with this requirement are applications such as verifiable blind quantum computing \cite{leichtle_verifying_2021} and quantum secret sharing \cite{crepeau2002secure}, as well as important subroutines such as entanglement purification \cite{bennett1996purification,deutsch1996quantum}.
In some contexts, multiple simultaneously-existing links are collectively referred to as an \textit{entanglement packet} \cite{beauchamp2025modular}. An entanglement packet is depicted in Figure \ref{fig:entanglement}(b).
The fast generation of entanglement packets is a task of fundamental importance for a functional quantum network. In this work, we find protocols that optimise the rate of entanglement packet generation, by adaptively varying a rate-fidelity trade-off mechanism available due to the entanglement generation scheme.

Here, we consider a setting with two nodes that attempt entanglement generation sequentially. 
In our model we assume that time is divided into discrete, uniform time steps, where in each time step, a single entanglement generation attempt is performed.
This is very often the case in near-term quantum networks, where heralded entanglement generation schemes succeed probabilistically and take up a fixed amount of time, due to the transfer of classical and quantum information between distant nodes~\cite{barrett_efficient_2005,cabrillo1999creation,duan2001long,sangouard2007long,beukers2024remote}. The true timescale of an attempt depends on several factors. For example, the distance between the two nodes determines the time taken for classical and quantum information transfer. Moreover, the duration of an attempt may depend on other quantum network subroutines such as the number of batched executions of physical entanglement generation scheme that defines a single attempt (see Section \ref{sec:exp_setup} for more details of batching).
To keep our model completely agnostic to the entanglement generation scheme being employed, we keep the time units abstract, and a single attempt is assumed to take up one unit of time.

After an attempt, an entangled link is generated with success probability $p$. 
The link is generated with initial fidelity ${F = \bra{\Psi_{00}}\rho\ket{\Psi_{00}}}$ to the target maximally-entangled state $$\ket{\Psi_{00}} = \frac{1}{\sqrt{2}}\left(\ket{00}+\ket{11}\right).$$ 
When a link is generated successfully, it is assumed to be immediately stored in memory. 
While the entangled link is stored in memory, it is subject to time-dependent noise (or \textit{decoherence}), which causes the fidelity to degrade over time. 
We assume a simple depolarising noise model: given a link with initial fidelity $F$, the fidelity of the link after $t$ time steps is 
\begin{equation}
    F \mapsto e^{-\Gamma t}\left(F - \frac{1}{4}\right) + \frac{1}{4},
    \label{eqn:decoherence}
\end{equation}
where $\Gamma$ is the decoherence rate. 
The depolarising noise model can be seen as a worst-case noise model, because any noise model can be brought to this form with a twirling map \cite{horodecki1999reduction}.  
An exponential decay in fidelity with time is a simple and commonly-used noise model for decoherence  \cite{humphreys2018deterministic,hermans2022qubit}.

We assume that there is a fixed minimum fidelity $F_\mathrm{app}$ required by an application for each link. A link is discarded once the fidelity of the link decays below $F_\mathrm{app}$  (see Figure \ref{fig:entanglement}). The goal is to generate $n$ simultaneously-existing links. Accordingly, we assume that each node has $n$ memories, each with the same decoherence rate $\Gamma$.

Suppose that at time $t=0$, there are no links stored in memory. We let $T$ denote the first time that there are $n$ simultaneously-existing links in memory. An important performance metric to consider is the expected time $\mathbb{E}[T]$ until completion, or alternatively the time until a terminal state is reached.  A terminal state is any state with $n$ links in memory. See Figure \ref{fig:entanglement} for a depiction of an intermediate state, with fewer than $n$ links in memory, and a terminal state. 
Depending on the system parameters, $\mathbb{E}[T]$ may be excessively large. For example, if the decoherence rate $\Gamma$ is large, then links are discarded soon after they are generated. 
Therefore, any link in memory is likely to be discarded before enough remaining links are successfully generated for the application. The same holds if the probability of generating a link is small, or if the initial fidelity of entangled states is small. 

We consider a setting where, before each entanglement generation attempt, the system may choose the generation parameters $(p,F)$ from a fixed set
\begin{equation}
    \mathcal{A} = \{(p_i,F_i):i = 1,\dots, |\mathcal{A}|\},
    \label{eqn:set_actions}
\end{equation}
where it is assumed that $F_i > F_{\mathrm{app}}$ for all $i$.
The choice of generation parameters at time $t$ may depend on the current state of the system $S_t$, which in our model is defined by the number of links that are in memory at time $t$ and their corresponding fidelities.

In many entanglement generation schemes, there exists a trade-off between the probability $p$ of successful entanglement generation and the fidelity $F$ of the generated link. In our model, this is captured by enforcing 
\begin{equation}
    F_i<F_j \Leftrightarrow p_i>p_j, \;\;\; i\neq j.
    \label{eqn:rate_fidelity_tradeoff_action}
\end{equation}
Due to the many possible entanglement generation schemes and ways that this trade-off can arise, we keep the set of $(p,F)$ pairs deliberately abstract as in (\ref{eqn:set_actions}) and (\ref{eqn:rate_fidelity_tradeoff_action}).

A well-known example of the trade-off between $p$ and $F$ arises in the single-click protocol \cite{cabrillo1999creation}, where varying the bright-state parameter $\alpha$ results in a linear relationship between the two, $(p,F) = (\kappa \alpha,1- \alpha) $, where $\kappa$ is a constant depending on hardware parameters such as the photon loss \cite{hermans_entangling_2023}. The trade-off between $p$ and $F$ exists in other physical entanglement generation schemes and could for example be due to the number of entanglement generation attempts that occur before the system is reinitialised \cite{pepsi}, or due to the mean photon number when a weak coherent pulse is used as a single-photon source \cite{knaut2024entanglement}. The same trade-off is also ubiquitous at a higher level: for example, if entanglement purification is employed, then in the choice of protocol there exists a fundamental trade-off between the output fidelity and the probability of the purification protocol succeeding \cite{rozpkedek2018optimizing}. We assume that the set of generation parameters (\ref{eqn:set_actions}) is finite without loss of generality (see Section \ref{sec:exp_setup} for details of how an infinite action set may be reduced to a finite one).

We note that certain platforms and/or entanglement generation schemes may not admit an adaptable trade-off. One example is the double-click entanglement scheme \cite{barrett_efficient_2005} that, in a noiseless scenario, generates an entangled pair with unit fidelity conditional on two photons arriving at a central heralding station. Within such a scheme, there is no mechanism to increase the success probability at the expense of the fidelity, and so the corresponding action set (\ref{eqn:set_actions}) contains only a single action. The analysis of the completion time $\mathbb{E}[T]$ then reduces to the constant-action scenario, which was analysed extensively in \cite{davies_tools_2024}. However, higher-level network protocols in a network based on such a platform may still admit an adaptable trade-off, such as in the choice of entanglement purification protocol, or the cut-off time applied in a quantum repeater chain.

In this work, we model the system as a Markov decision process (MDP), which we subsequently solve to find optimal policies. A policy $\pi$ is a map\footnote{In general, a policy can be non-deterministic \cite{sutton_reinforcement_markov}. However, since the state set is finite in our problem, there always exists a deterministic optimal policy \cite{szepesvari2022algorithms}. For clarity, we therefore introduce the policy as deterministic.}
\begin{equation}
    \pi: \mathcal{S} \rightarrow \mathcal{A},
    \label{eqn:policy_def}
\end{equation}
where $\mathcal{A}$ is the set of generation parameters (\ref{eqn:set_actions}) (also referred to as the set of \textit{actions}) and $\mathcal{S}$ is the state set. When implementing policy $\pi$, the generation parameters chosen at time $t$ depend on the current state $S_t$ and are given by $\pi(S_t) \in \mathcal{A}$.
The state $S_t$ and the action $\pi(S_t)$ determine the possible values of the state in the next time step, $S_{t+1}$. 
When the policy $\pi$ is employed, we write the completion time as $T \equiv T_{\pi}$. The policy $\pi^*$ is called \textit{optimal} if it minimises the expected completion time, 
\begin{equation}
    \mathbb{E}[T_{\pi^*}] = \min_{\pi} \mathbb{E}[T_{\pi}].
\end{equation}

Now that we have introduced the problem, we outline our main contributions:
\begin{itemize}
    \item We firstly ask and answer the question: ``\textit{what is the maximum speed-up achievable with adaptivity?}''. We use dynamic programming \cite{sutton_reinforcement_markov} to compute optimal policies $\pi^*$ and the optimal performance $\mathbb{E}[T_{\pi^*}]$. We compare the optimal performance with the performance of the best constant-action policy $\pi_{\mathrm{con}}$ that chooses the same action in every time step. We evaluate policy performance in two parameter regimes that we call the \textit{near-term regime} and the \textit{far-term regime}. The parameters of the near-term regime are based on recent experiments. The far-term regime is assumed to have an improved memory lifetime, and therefore has a more complex state set and policy behaviour. Both parameter regimes have an action set (\ref{eqn:set_actions}) given by the trade-off relation from the single-click protocol \cite{cabrillo1999creation}. In both parameter regimes, we find that the optimal policy can provide a speed-up over the constant-action policy of as much as 
    \begin{equation}
        \frac{\mathbb{E}[T_{\pi^*}]}{\mathbb{E}[T_{\pi_{\mathrm{con}}}]} \approx 0.05.
        \label{eqn:approximate_speedup_ratio}
    \end{equation}
    We conclude that, given an adjustable rate-fidelity trade-off available in the entanglement generation scheme, it can be highly advantageous to use adaptive protocols to boost the generation rate of entanglement packets. Moreover, we see that this advantage increases with $n$, indicating that adaptive polices will provide even more improvement as quantum networks become more sophisticated.
    \item We secondly ask and answer the question: ``\textit{is a comparable speed-up achievable with simple heuristic policies?}''. By computing $\pi^*$ in the two parameter regimes, we gain insights about the structure of optimal policies. Based on our insights, we define a simple heuristic policy $\pi_{\mathrm{h}}$. 
    Remarkably, we find that in the near-term regime, the heuristic policy is in fact optimal, i.e. $\mathbb{E}[T_{\pi_{\mathrm{h}}}] = \mathbb{E}[T_{\pi^*}]$. In the far-term regime, the heuristic policy performs close to optimally, satisfying
    \begin{equation}
         \frac{\mathbb{E}[T_{\pi_{\mathrm{h}}}] - \mathbb{E}[T_{\pi^*}]}{\mathbb{E}[T_{\pi^*}]} < 0.03
    \end{equation}
    for all $n$ for which $\pi^*$ was computed. 
    Moreover, we verify empirically that $\pi_{\mathrm{h}}$ is computable in parameter regimes where $\pi^*$ is not. 
    In parameter regimes where it is not possible to compute an optimal policy $\pi^*$ due to scaling of the state set $|\mathcal{S}|$, one may therefore instead employ $\pi_{\mathrm{h}}$ and expect either optimal or close-to-optimal performance. For such a case in our far-term regime, the heuristic policy provides a speed-up over the constant-action policy of as much as  
    \begin{equation}
        \frac{\mathbb{E}[T_{\pi_\mathrm{h}}]}{\mathbb{E}[T_{\pi_{\mathrm{con}}}]}\approx 1.05\cdot 10^{-6}.
    \end{equation}
\end{itemize}
The remainder of this work is structured as follows. In Section~\ref{sec:related_work}, we summarise related work. Then, in Section~\ref{sec:methods}, we formally define the MDP and briefly introduce dynamic programming. In Section~\ref{sec:results}, we present our results: in Section~\ref{sec:analysis_n=2} we firstly present an analytical solution for the optimal policy and its performance for $n=2$. In Section~\ref{sec:heuristic}, we define the heuristic policy. The MDP, analytical solution for $n=2$ and heuristic policy are all introduced for a generic state set and generic action set (entanglement generation scheme). In Section~\ref{sec:performance_comparison} we present our numerical results. We consider the example of a single-click entanglement generation scheme, and compare the performance of the optimal policy, heuristic policy and baselines in the two parameter regimes of interest. We also extract general conclusions about the properties one can expect of optimal policies. We also analyse how the performance and structural properties change with different values of $F_\mathrm{app}$, the minimum required fidelity.
Finally, we conclude and suggest possible future extensions of our work in Section~\ref{sec:discussion}.

\section{Related work}
\label{sec:related_work}
The scenario studied in this work is a generalisation of the one studied in \cite{davies_tools_2024}. In \cite{davies_tools_2024}, the generation parameters $(p,F)$ were assumed to be the same in each time step. In this work, we allow the system to choose instead from the set of generation parameters (\ref{eqn:set_actions}). The system studied in \cite{davies_tools_2024} is therefore equivalent to the constant-action policy $\pi_{\mathrm{con}}$ that chooses the same generation parameters in every state.
We note that the methods used in this work are very different from \cite{davies_tools_2024}: here we formulate the problem as an MDP and perform optimisation with dynamic programming, whereas in \cite{davies_tools_2024} analytical solutions were derived for $\mathbb{E}[T_{\pi_\mathrm{con}}]$. 

MDP-based techniques have previously shown their value for the optimisation of a range of quantum network protocols. For example, they have been used to find optimal entanglement swapping policies in repeater networks  \cite{khatri_policies_2021,inesta_optimal_2023, khatri_design_2022}. Approximate reinforcement learning approaches, which efficiently find approximate solutions to problems formulated as MDPs, have also been utilised in the context of quantum networks, for example for designing entanglement routing schemes \cite{le_entanglement_2022}, optimising quantum repeater chains for secret key distribution \cite{reis_deep_2023} or entanglement distribution \cite{haldar_fast_2024}, and designing new and improved communication protocols, particularly in networks with asymmetric features \cite{wallnofer_machine_2020}. We also note that other works have optimised the performance of quantum network protocols by varying the trade-off between the success probability and output fidelity, most often optimising over the bright-state parameter \cite{qnum,avis_requirements_2023,avis2025stochastic}. 
However, other than \cite{davies_tools_2024}, the aforementioned studies all optimise the delivery of a single entangled link. In our work, we optimise the delivery of multiple links in the form of an entanglement packet, which is a fundamentally different problem. 

\section{Methods}
\label{sec:methods}

\subsection{Constructing the Markov decision process}
\label{sec:mdp_construction}
An MDP is defined as a 5-tuple $(\mathcal{S}^+, \mathcal{A}, \mathrm{P}, R, \mathcal{S}_n)$ where $\mathcal{S}^+$ is the state set, $\mathcal{A}$ is the action set, $\mathrm{P}$ is the transition function, $R$ is the reward function, and $\mathcal{S}_n\subset \mathcal{S}^+$ is the set of terminal states. We will also use the notation $\mathcal{S}\coloneqq \mathcal{S}^+ \setminus \mathcal{S}_n$ to denote the state set excluding the terminal states. 
In the following, we elaborate on each component of the MDP for our system.
In Appendix \ref{app:timing}, we provide a detailed breakdown of the events assumed to occur within each time step, upon which the transitions of our MDP are based.
From now on, we assume arbitrary time units such that a single time step has $\Delta t = 1$.
\subsubsection{State set}
\label{subsec:state_space}
The fidelity of each link in memory is fully characterised by the time it will survive before being discarded, which we refer to as the \textit{time-to-live} (TTL) of a link. 
Suppose that a link has fidelity $F>F_\mathrm{app}$. Recalling our decoherence model (\ref{eqn:decoherence}), the TTL of the link is given by
 \begin{equation}
    t_{\mathrm{TTL}}(F) = \ceil*{\frac{1}{\Gamma}\ln{\frac{F - \frac{1}{4}}{F_\mathrm{app} - \frac{1}{4}}}}.
    \label{eqn:TTL}
\end{equation}
The ceiling function is taken because we work in discrete time. Letting the maximum fidelity of a newly generated link be denoted by $$F_{\max} = \max \{F : (p,F) \in \mathcal{A} \},$$ the maximum TTL is given by
\begin{equation}
    t_{\max} = t_{\mathrm{TTL}}(F_{\max}).
    \label{eqn:t_max}
\end{equation}
The state $S_t$ of the system at time $t$ characterises the relevant system information. For our system, the state is the number of links stored in memory and their TTLs (corresponding to link fidelities). If there are $m$ links in memory, the state $s$ is given by
\begin{equation}
   s =  \{t_1,\dots, t_m \}, 
   \label{eqn:example_state}
\end{equation}
where $t_i\in [t_{\max}]$ is the TTL of the $i$th link, and $[t_{\max}] = \{ 1,\dots, t_{\max}\}$. Formally, $s$ is a multiset that may contain multiple elements of the same value, because it is possible that $t_i = t_j$ for $i\neq j$. Since the $n$ memories are assumed to be identical, the ordering of the TTLs in any given state is assumed to be decreasing without loss of generality, i.e. $t_i \geq t_j$ for $i<j$.

As an example, if $S_t=\{3,2\}$ then at time $t$ there are two links in memory, one of which will be discarded after three time steps, and the other after two time steps. The first link has a higher fidelity than the second link, because it will be discarded later.

We denote the set of all states with $m$ links in memory as
\begin{equation}
    \label{eq:state}
    \mathcal{S}_m = \big\{\left\{t_1,\dots, t_m \right\} : t_i \in [t_{\mathrm{max}}] , t_1 \geq \dots \geq t_m \big\}.
\end{equation}
We denote the state with no links in memory as the empty set $\varnothing \equiv \{ \}$. We study the MDP in an episodic setting, where completion is achieved when there are $n$ links in memory, or equivalently, when it reaches a state $s \in \mathcal{S}_n$. 
See Figure \ref{fig:entanglement} for an illustration.
We then define
\begin{equation}
    \label{eqn:states}
    \mathcal{S} =  \bigcup_{m=1}^{n-1} \mathcal{S}_m \cup \{\varnothing\},
\end{equation}
so that
\begin{equation}
    \mathcal{S}^+ = \mathcal{S} \cup \mathcal{S}_n.
\end{equation}
The time $T_{\pi}$ until completion (when policy $\pi$ is employed) may then be written explicitly as 
\begin{equation}
    T_{\pi} = \min\{t: S_t \in \mathcal{S}_n\}.
\end{equation}
\subsubsection{Action set}
\label{subsec:action_space}
The action set $\mathcal{A}$ is the set of generation parameters (\ref{eqn:set_actions}). The possible generation parameters depend on the specific entanglement generation scheme used. We note that $\mathcal{A}$ is without loss of generality finite (see Section~\ref{sec:performance_comparison}). In Section~\ref{sec:performance_comparison} we give an example of $\mathcal{A}$ for the single-click protocol.
\subsubsection{Transition function}
\label{subsec:transition_function}
We now define explicitly the transition function $\mathrm{P} : \mathcal{S} \times \mathcal{A} \rightarrow \Delta(\mathcal{S}^+)$, where $\Delta(\mathcal{S}^+)$ denotes the set of all probability distributions over $\mathcal{S}^+$.
Suppose that at time $t$, the system is in state $s$, and that action $a\in \mathcal{A}$ is chosen. Then, the transition function $\mathrm{P}(s'|s,a)$ determines the probability of transitioning to state $s'$ in time step $t+1$. We now write down the transition function  $
\mathrm{P}$ explicitly for our system.
We suppose that $s = \{t_1,\dots, t_m \}$ and that the action chosen is $a = (p,F)$. There are two transitions that can occur. The first is when the entanglement generation attempt succeeds, which occurs with probability $p$. In this case, a new link is generated with TTL $t_{\mathrm{TTL}}(F)$, as given by (\ref{eqn:TTL}). All other links decohere by one time step and are discarded if the fidelity falls below $F_{\mathrm{app}}$, or equivalently when the TTL becomes zero. Since the TTL (\ref{eqn:TTL}) is the inverse of the decoherence map (\ref{eqn:decoherence}), decoherence over a single time step simply causes all TTLs to reduce by one.
In the event of entanglement generation success, the state in the next time step is therefore given by
\begin{equation}
    s'_{\mathrm{succ}} =  \{t_j-1 : t_j \in s, t_j>1 \} \cup \{t_{\mathrm{TTL}}(F)\}.
\end{equation}
The second possible transition is when entanglement generation fails, which occurs with probability $1-p$. The state in the next time step is given by
\begin{equation}
    s'_{\mathrm{fail}} =  \{t_j-1 : t_j \in s, t_j>1 \}.
\end{equation}
We therefore have 
\begin{equation}
    \mathrm{P}(s'|s,a) = 
    \begin{cases}
         p, &\text{ if } s' = s'_{\mathrm{succ}} \\
        1-p, &\text{ if } s' = s'_{\mathrm{fail}} \\ 
        0, &\text{ otherwise.}
    \end{cases}
    \label{eqn:transition_fn_1}
\end{equation}
Similarly, the transitions from the state $\varnothing$ are given by
\begin{equation}
    \mathrm{P}(s'|\varnothing,a) = 
    \begin{cases}
         p, &\text{ if } s' =  \{t_{\mathrm{TTL}}(F)\} \\
        1-p, &\text{ if } s' = \varnothing \\ 
        0, &\text{ otherwise.}
    \end{cases}
    \label{eqn:transition_fn_2}
\end{equation}
The transition function is fully defined by (\ref{eqn:transition_fn_1}) and (\ref{eqn:transition_fn_2}).

The transition function is particularly simple in our case due to the assumption (\ref{eqn:decoherence}) that fidelity decays exponentially in time, meaning that the TTL of each link will simply decrease by one as we advance one time step. If noise is no longer Markovian, then it would be necessary to redefine the state set and transition function. We leave such a study for future work.

\subsubsection{Reward}
Since our objective is to minimise the expected completion time $\mathbb{E}[T_{\pi}]$, the reward is $R(s,a) =  -1$ for all $a\in \mathcal{A}$ and $s\in \mathcal{S}$.
\subsection{Dynamic programming}
\label{sec:policy_iteration}
We use a dynamic programming algorithm known as \textit{policy iteration} to compute optimal policies. We let $R_t \coloneqq R(S_t,a)$ be the reward at time step $t$, given that the state is $S_t$ and action $a$ is taken. Then, the \textit{value} of a state $s\in\mathcal{S}$ under policy $\pi$ is defined as 
\begin{equation}
    V_{\pi}(s) \coloneqq  \mathbb{E}\left[ \sum_{k=0}^{T_{\pi}-t-1} R_{t+k} \Bigg| S_t = s\right].
\end{equation}
This is the definition value for episodic problems with no discounting factor -- see e.g. Chapter 3 of \cite{sutton_reinforcement_markov} for more details of how the value is defined for other systems.
In our system, 
\begin{align}
    V_{\pi}(s) &=  \mathbb{E}\left[ \sum_{k=0}^{T_{\pi}-t-1} -1 | S_t = s\right] \\
    &= -\mathbb{E}[T_{\pi}-t|S_t = s]
    \\ &= -\mathbb{E}[T_{\pi}|S_0 = s].
\end{align}
Moreover, $V_{\pi}(s)=0$ for all $\pi$ and $s\in \mathcal{S}_n$.
In the final step, we have used the fact that our MDP is time-homogeneous.
Then, $V_{\pi}(s)$ is simply the expected completion time, given that the process starts in state $s$ and policy $\pi$ is employed.

For all $s\in \mathcal{S}$ and any policy $\pi$, we can calculate $V_{\pi}(s)$, which is a step known as \textit{policy evaluation}. We do this by solving a linear system known as the \textit{Bellman equation} \cite{sutton_reinforcement_markov}, which is given by
\begin{equation}
    \label{eq:bellman}
    V_\pi (s) \coloneqq -1 + \sum_{s' \in \mathcal{S}^+} \mathrm{P}(s' | s, \pi(s)) V_\pi (s').
\end{equation}
Policy iteration then computes an optimal policy $\pi^*$ by starting with an arbitrary policy $\pi_0$, and then iteratively updating it until the value $V_{\pi}(s)$ of each state $s$ has been maximised.
Let $\pi_k$ be the policy at the $k$-th iteration. 
The policy is then updated by maximising the value with respect to all $a\in\mathcal{A}$ for each state, which again can be achieved by considering the Bellman equation for $V_{\pi_{k-1}}$,
\begin{align}
\pi_{k}(s) &= \underset{a \in \mathcal{A}}{\mathrm{argmax}\,}\sum_{s'\in \mathcal{S}^+}\mathrm{P}(s'|s,a)(-1+V_{\pi_{k-1}}(s')) \\ &= \underset{a\in \mathcal{A}}{\mathrm{argmax}\,}\sum_{s'\in \mathcal{S}^+}\mathrm{P}(s'|s,a)V_{\pi_{k-1}}(s').
\end{align}
Since both $|\mathcal{S}|$ and $|\mathcal{A}|$ are finite in our MDP, policy iteration converges to an optimal policy $\pi^*$ after a finite number of iterations \cite{sutton_reinforcement_markov}.

\section{Results}
\label{sec:results}
\subsection{Analytical solution for $n=2$}
\label{sec:analysis_n=2}
We now analyse the case $n=2$, for which we are able to find a closed-form expression for the optimal policy $\pi^*$ and its expected waiting time. This will help to provide intuition for the properties we expect from optimal policies for larger $n$. In Section \ref{sec:heuristic}, we use this intuition to define a heuristic policy for $n>2$ that we later see displays near-optimal performance, while remaining efficient to compute. 

We recall the definition of the state set (\ref{eqn:states}).
For $n=2$, the state set is written explicitly as
\begin{equation}
    \mathcal{S} = \left\{\varnothing, \{1\},\{2\},\dots,\{t_{\max}\} \right\}.
\end{equation}
For now, we write $\pi^*(\varnothing) = (p_{\varnothing},F_{\varnothing}) \in \mathcal{A}$. The precise choice of $\pi^*(\varnothing)$ will be fixed later.
We then notice that, if a single link is present in memory, one must maximise the success probability. This is because the process will be completed if the second link is generated before the first is discarded. Then, although maximising the success probability also minimises the fidelity of the generated link,  this would not matter because the second link does not impact future behaviour. In particular, we define the action
\begin{align}
    (p_{\max},F_{ \min})=\underset{(p,F)\in \mathcal{A}}{\text{argmax}}\{p\}.
    \label{eqn:max_prob_action}
\end{align}
Note that by the rate-fidelity trade-off (\ref{eqn:rate_fidelity_tradeoff_action}) assumed in $\mathcal{A}$, we also have $F_{ \min} = \min \{F : (p,F)\in \mathcal{A}\}.$ We then set 
\begin{equation}
    \pi^*(s) \coloneqq (p_{\max},F_{ \min}), \text{  for } s\in \mathcal{S}_1 \setminus \big\{ \{ 1 \}\big\}.
    \label{eqn:opt_policy_n=2_maxprob}
\end{equation}
In (\ref{eqn:opt_policy_n=2_maxprob}), we have not assigned the same action to $\pi^* (\{1 \})$ because, when there is a link with TTL of one, the link will be discarded in the next time step. Thus, there is no chance of completing the process in the next time step. One should therefore not maximise the success probability, as we do for the other states in (\ref{eqn:opt_policy_n=2_maxprob}). In particular, $\varnothing$ and $\{1\}$ both contain zero \textit{viable links}, which are the number of links that have a non-zero probability of surviving until the system reaches a terminal state $s\in \mathcal{S}_n$. We therefore view $\varnothing$ and $\{1\}$ as equivalent states, and they are assigned the same action
\begin{equation}
    \pi^* (\{1 \}) = \pi^*(\varnothing) = (p_{\varnothing},F_{\varnothing}).
    \label{eqn:opt_policy_n=2_emptystate}
\end{equation}
Equivalent states may also be identified when $n>2$ to perform a reduction of the state set $\mathcal{S}$ (see Appendix \ref{sec:statespacesize}), which can speed up the computation of optimal policies with dynamic programming.

Given the structure of the optimal policy from (\ref{eqn:opt_policy_n=2_maxprob}) and (\ref{eqn:opt_policy_n=2_emptystate}), we now 
explicitly compute the expected completion time $ \mathbb{E}[T_{\pi^*}]$. Assuming that the system starts in the state $S_0 = \varnothing$, it will alternate between two phases until completion:
\begin{enumerate}[label=(\Alph*)]
    \item There are no viable links in memory (the state is either $\varnothing$ or $\{1\}$). The system attempts entanglement generation with parameters $(p_{\varnothing},F_{\varnothing})$ until a link is successfully generated.
    \item The newly generated link survives in memory for $t_{\mathrm{TTL}}(F_{\varnothing})$ time steps, until it is discarded. During the first $t_{\mathrm{TTL}}(F_{\varnothing}) - 1$ time steps when the link is in memory, the system attempts entanglement generation with parameters $(p_{\max},F_{ \min})$. 
    If at least one of these attempts is successful, the process is completed.
    If none of these attempts are successful, then in the final time step, the state is $\{1\}$ and the system returns to phase (A).
\end{enumerate}
The change in action of the optimal policy as the system moves from phase (A) into phase (B) captures the adaptivity of the optimal policy: a constant-action policy would choose the same action in all states. 

The expected completion time may then be computed using exactly the same method as the one used in Section III-B of \cite{davies_tools_2024}. The method makes use of properties of the geometric distribution. We obtain
\begin{equation}
        \mathbb{E}[T_{\pi^*}]=\frac{1}{p_{\max}} +  \frac{1}{p_{\varnothing}\left(1 - (1 - p_{\max})^{t_{\mathrm{TTL}}(F_\varnothing) - 1}\right)},
        \label{eq:2action2links-expectation}
\end{equation}
and the solution for $\pi^*(\varnothing)$ and $\pi^*(\{1\})$ is therefore given by 
\begin{equation}
    (p_{\varnothing},F_{\varnothing}) = \underset{(p,F)\in \mathcal{A}}{\mathrm{argmin}} \left\{\frac{1}{p\left(1 - (1 - p_{\max})^{t_{\mathrm{TTL}}(F) - 1}\right)} \right\}.
    \label{eqn:formula_action_emptystate}
\end{equation}
We have now fully defined the optimal policy $\pi^*(s)$ for all $s\in \mathcal{S}$, and this completes the analytical solution. We have also derived a formula for its performance in (\ref{eq:2action2links-expectation}).

The analysis for $n=2$ offers valuable insights into the expected behaviour of the optimal policy for $n>2$.
As we extend to cases with larger $n$, we anticipate that the optimal policy will continue to select the action $(p_{\max},F_{ \min})$ for states with $n-1$ viable links in memory. For other states, the action chosen by the optimal policy must correctly balance the probability of generation $p$ with the time-to-live $t_{\mathrm{TTL}}(F)$. This is because a link must be generated quickly, but there must also be sufficient time for the remaining links to be generated while that link is in memory. For $n=2$, the correct balance is captured by (\ref{eqn:formula_action_emptystate}).

\subsection{Heuristic policy for $n>2$}
\label{sec:heuristic}
Drawing on the intuition acquired with the derivation of the optimal policy for $n=2$ in Section \ref{sec:analysis_n=2}, we now present a heuristic policy $\pi_{\mathrm{h}}$. 
The heuristic policy has the properties that we intuitively expect from good policies, as discussed in the previous section for $n=2$: a good policy should start with high-fidelity choices of action and boost the probability as the system approaches completion. We note that there are many ways in which one could define a heuristic with these properties, and the one presented here is just a single example. In this way, the $n=2$ heuristic is not claimed to be a general structural foundation for the $n>2$ heuristic, but a motivation.

In our experiments (Section \ref{sec:performance_comparison}), we verify empirically that the heuristic is efficient to compute for large $n$. Later, we find that our heuristic performs either optimally or close to optimally in the parameter regimes explored numerically.

The size of the state set $|\mathcal{S}|$ scales exponentially with $n$, meaning that for large $n$ computing optimal policies with policy iteration (Section \ref{sec:policy_iteration}) is intractable. See Appendix~\ref{sec:statespacesize} for a detailed analysis of the scaling. With our setup, we are only able to compute optimal policies for $n\leq 7$ in the parameter regimes that we explore in Section~\ref{sec:performance_comparison}. 
Here, we find that our heuristic policy $\pi_\mathrm{h}$ is efficient to compute for large $n$ and performs close to optimally for $n\leq 7$ (see Section \ref{sec:performance_comparison}). 
One can therefore expect $\pi_{\mathrm{h}}$ to exhibit high performance for $n>7$.

We now define $\pi_\mathrm{h}$. Recalling the notion of a viable link that was introduced in \autoref{sec:analysis_n=2}, we define the function $$N_v: \mathcal{S} \rightarrow \{0,1,\dots,n-1 \}$$ such that $N_v(s)$ outputs the number of viable links in state $s$. For example, recalling the discussion in \autoref{sec:analysis_n=2}, we have $N_v(\varnothing) = N_v(\{1\})=0$. More generally, given the state $s = \{t_1,\dots ,t_m \}$, the $m$-th link with TTL $t_m$ is viable if there is a chance of completing the process (i.e. generating $n-m$ remaining links) before the link expires. Therefore, the link is viable if $t_m > n-m$. Recall from Section \ref{subsec:state_space} that the labelling of the TTLs is assumed to be decreasing, i.e. $t_i \geq t_j$ for $i<j$. If the $m$-th link is viable, then for all $k \in \{1,\dots, m-1 \}$, we therefore have $t_k> n-m$ and all other links are viable. The total number of viable links in the state $s$ is given by
\begin{equation}
    N_v(s) \coloneqq \max \{j: t_j>n-j \}.
    \label{eqn:N_v_definition}
\end{equation}
Given the state $s = \{t_1,\dots, t_m \}$, we define a second function $v : \mathcal{S} \rightarrow \mathcal{S}$ such that
\begin{equation}
    v(s) = \begin{cases}
        \{t_1,\dots,t_{N_v(s)}\} &\text{  if  } N_{v}(s) \geq 1
        \\ \varnothing, &\text{  if  } N_{v}(s) = 0.
    \end{cases}
    \label{eqn:v_definition}
\end{equation}
The state $v(s)$ contains the viable links of $s$. Identifying ${s\equiv v(s)}$ can be used to reduce the size of the state set ---  see Appendix \ref{sec:statespacesize} for more details. Although we do not implement this in our numerical experiments, one could adapt the MDP that was defined in Section \ref{sec:methods} to the reduced state set and find a a speed-up in the computation of optimal polices. 

We now define the heuristic policy $\pi_{\mathrm{h}}$. As we did in Section~\ref{sec:analysis_n=2} for the case $n=2$, we set the actions for states with zero viable links to an arbitary value,
\begin{equation}
    \pi_{\mathrm{h}}(s) \coloneqq (p_{\varnothing},F_{\varnothing}) \in \mathcal{A} \text{,    if    } N_{v}(s) = 0.
    \label{eqn:heuristic_policy_zero_viable}
\end{equation}
Further, we use the intuition established in Section \ref{sec:analysis_n=2} and enforce that the policy must choose the maximum-probability action (\ref{eqn:max_prob_action}) when there are $n-1$ viable links in memory,
\begin{equation}
    \pi_{\mathrm{h}}(s) \coloneqq (p_{\max},F_{ \min}), \text{    if    } N_v(s) = n-1. \label{eqn:heuristic_policy_value_n-1_links}
\end{equation}
We now consider states $s$ such that $0< N_v(s)< n-1 $. 
For such states, the heuristic policy chooses the highest-probability action that generates a link TTL greater than or equal to the smallest TTL of a viable link in memory. Explicitly, we set
\begin{align}
    \pi_{\mathrm{h}}(s) \coloneqq \underset{(p,F) \in \mathcal{A} }{\mathrm{argmax}} \big\{ p: \;  t_{\mathrm{TTL}}&(F) \geq t_{N_v(s)} - 1\big\}, \label{eqn:heuristic_policy_value}  \\ &\text{if    } 0< N_v(s)< n-1 . \nonumber
\end{align}
In particular, we see that the action taken in a given state $s$ is only dependent on its viable links $v(s)$,
\begin{equation}
    \pi_{\mathrm{h}}(s) = \pi_{\mathrm{h}}(v(s)) \text{  for all  } s\in \mathcal{S}.
\end{equation}
The definitions (\ref{eqn:heuristic_policy_value_n-1_links}) and (\ref{eqn:heuristic_policy_value}) capture the adaptivity of the heuristic policy. From (\ref{eqn:heuristic_policy_value}), we see that when the TTL of the links in memory is smaller, the policy chooses an action with a larger success probability. When there are $n-1$ links in memory, the heuristic policy adapts again via (\ref{eqn:heuristic_policy_value_n-1_links}) to choose the maximum success probability action. 

In (\ref{eqn:heuristic_policy_value_n-1_links}) and (\ref{eqn:heuristic_policy_value}), we have now fixed all actions apart from $(p_{\varnothing},F_{\varnothing})$, from (\ref{eqn:heuristic_policy_zero_viable}). The choice of $(p_{\varnothing},F_{\varnothing})$ may be fixed by either policy evaluation or performing a Monte Carlo simulation of the system while employing $\pi_{\mathrm{h}}$ with each $(p_{\varnothing},F_{\varnothing})\in \mathcal{A}$. Then, one may select the value that minimises $\mathbb{E}[T_{\pi_{\mathrm{h}}}]$. This step involves a maximum of $|\mathcal{A}|$ policy evaluation steps, either carried out exactly by solving the Bellman equations (\ref{eq:bellman}), or approximately by performing a Monte Carlo simulation. The possibility to evaluate performance with simulation allows for the computation of the heuristic and its performance for regimes with larger state sets than the optimal policy (see Section \ref{sec:results}). As $n$ increases, the computation of the heuristic is therefore only limited by the Monte Carlo simulation of $|\mathcal{A}|$ different policies. Thus, if the Monte Carlo simulation is feasible, then so is the computation of the heuristic. In our numerical experiments in Section \ref{sec:performance_comparison}, we find that this enables the computation of $\pi_\mathrm{h}$ in regimes where $\pi^*$ cannot be computed with policy iteration.

The policy $\pi_{\mathrm{h}}$ is particularly simple in the case where, for any $t\in \{ 1,\dots, t_{\max}\}$, there exists $(p,F)\in\mathcal{A}$ such that $t_{\mathrm{TTL}}(F) = t$. In other words, in $\mathcal{A}$ there are generation parameters that can produce a link with any $\mathrm{TTL}$. In such a case, (\ref{eqn:heuristic_policy_value}) is simply a \textit{matching heuristic}, which ensures that all viable links have the same $\mathrm{TTL}$, while maximising the success probability as much as possible.

The intuition behind our heuristic policy is as follows. With (\ref{eqn:heuristic_policy_value}) the heuristic policy ensures that, when at least one viable link in memory will soon expire (small TTL), the heuristic policy tries to quickly generate all links within the remaining time. On the other hand, if all links in memory have a large TTL, the heuristic policy tries to generate similarly long-lasting links. This is a property we will also see in Section \ref{sec:results} in the structure of optimal policies: it is a trend that, as links in memory decohere, the optimal policies prioritise higher-probability generation parameters. In fact, in one of the two parameter regimes considered, we see that the heuristic policy is in fact optimal.

It would also be interesting to analyse different classes of heuristic policies that do not follow the matching heuristic, as we consider here. For example, one could consider a binary policy, which employs an initial action when fewer than $k$ links are in memory, and a different action when there are $k$ or more links in memory. However, since our heuristic already performs extremely well, we leave such an investigation for future work.

\subsection{Performance comparison}
\label{sec:performance_comparison}
\subsubsection{Baselines}
\label{sec:baselines}
In the following, we compare our optimal and heuristic policies with two baseline policies. Firstly, we consider the \textit{best constant-action policy} $\pi_{\mathrm{con}}$, where
\begin{equation}
    \pi_{\mathrm{con}}(s) \coloneqq (p,F) \text{    for all    } s \in \mathcal{S}
\end{equation}
and $(p,F) \in \mathcal{A}$ is the value that minimises the performance $\mathbb{E}[T_{\pi_{\mathrm{con}}}]$.
The performance of this policy is well-studied with analytical methods in \cite{davies_tools_2024}.

As a second baseline, we also consider the policy $\pi_{\mathrm{ran}}$ with uniformly random actions. Unlike all  policies mentioned previously, this policy is non-deterministic. At each time step, the action is chosen from $\mathcal{A}$ uniformly at random:
\begin{align}
    \pi_{\mathrm{ran}}(S_t) = (p,F) &\text{ with prob. }\frac{1}{|\mathcal{A}|},\nonumber \\ &\text{ for all }(p,F)\in \mathcal{A}, \;\; t\geq 0.
\end{align}

\subsubsection{Parameter regimes}
\label{sec:exp_setup}
In the performance evaluation, we consider generation parameters that correspond to batched executions of the single-click protocol.
When photon loss is high, one execution of the single-click protocol has a very low success probability of generating entanglement \cite{hermans_entangling_2023}. Therefore, it can be beneficial for a single entanglement generation attempt to consist of $M$ executions of the single-click protocol, which increases the probability of success while minimising overhead due to communication with higher layers of the software stack \cite{pompili_experimental_2022}. We also refer to these as \textit{batched attempts}. The attempt is declared successful if at least one of the $M$ single-click executions succeeds.
In Appendix \ref{sec:tradeoff}, we show that for batched single-click attempts, the trade-off between $p$ and $F$ is accurately approximated as 
\begin{equation}\label{eq:trade-off}
    F = \lambda \ln(1-p) + 1.
\end{equation}
Here, $\lambda = 1/(2p_{\mathrm{det}}M)$ is a fixed parameter that depends on the probability of detecting an emitted photon $p_{\mathrm{det}}$ and the batch size $M$. 
The relation (\ref{eq:trade-off}) holds when $M \sim p_{\mathrm{det}}^{-1}$ is large.
The trade-off (\ref{eq:trade-off}) is valid for a sufficiently small range $p\in (0,q]$. Given that all newly generated links must have fidelity $F$ such that $F>F_{\mathrm{app}}$, by (\ref{eq:trade-off}) the maximum success probability $q$ satisfies
\begin{equation}
    q < 1-e^{\frac{F_\mathrm{app}-1}{\lambda}}.
\end{equation}
Noting that (\ref{eq:trade-off})  enables a continuous choice of $(p,F)$, we discretise the action set as follows. The maximum TTL of a newly generated link given the trade-off (\ref{eq:trade-off}) is given by 
\begin{equation}
    t_{\max} = \lim_{p\rightarrow 0}t_{\mathrm{TTL}}(\lambda \ln(1-p) + 1) = t_{\mathrm{TTL}}(1).
    \label{eqn:t_max_expts}
\end{equation}
Similarly, we let
\begin{equation}
    t_{\min} = t_{\mathrm{TTL}}( \lambda \ln (1-q) +1)
\end{equation}
be the minimum TTL of a newly generated link.
Then, for each  $i\in \{t_{\min},\dots,t_{\max} \}$, we define $p_i$ as the maximum probability with which one can generate a link with TTL $i$,
\begin{equation}
    p_i = \max\{p : t_{\mathrm{TTL}}(\lambda \ln(1-p) + 1) = i\}.
\end{equation}
We then work with the finite action set
\begin{equation}
    \mathcal{A} = \left\{(p_i, \lambda \ln(1-p_i) + 1) : i \in \{t_{\min},\dots ,t_{\max} \} \right\}.
    \label{eqn:actions_batch_single_click}
\end{equation}

We now compare the performance of our optimal and heuristic policies to our baselines in two parameter regimes. The first parameter regime has a high decoherence rate ${\Gamma = 0.19}$, which we refer to as the \textit{near-term regime}. The second parameter regime has a lower decoherence rate $\Gamma = 0.1$, which we refer to as the \textit{far-term regime}. Firstly, for both parameter regimes, we set the minimum required fidelity to be $F_{\mathrm{app}}=1/2$ and analyse in detail the behaviour of policies and their performance in Sections \ref{sec:near_term} and \ref{sec:far_sterm}. Then, in Section \ref{sec:changeFapp}, we consider the far-term regime and study how the performance and features of the optimal policies change for different values of $F_{\mathrm{app}}$. 

We note the value $F_{\mathrm{app}}=1/2$ is significant because it is the threshold value above which an isotropic state is entangled (see e.g. Appendix E of \cite{davies20241g1b}). Relatedly, certain well-known entanglement purification protocols are shown to be functional for noisy entangled states with a fidelity greater than $1/2$ \cite{bennett1996purification,deutsch1996quantum}. Furthermore, although $F_{\mathrm{app}}=1/2$ may be too low for some quantum applications to be executed, it is necessary to provide meaningful results for current parameter regimes (our near-term regime). As $F_{\mathrm{app}}$ increases in the near-term regime, it is quickly no longer feasible to generate entanglement packets with $n\geq 2$.

The parameter regimes are chosen as follows. Suppose that the memory lifetime of each memory qubit consists of $N$ executions of the single-click protocol. In recent experiments, $N\approx 5300$ \cite{hermans2022qubit}. Since an entangled link is stored in two qubits and decoherence acts on both of them, the total memory lifetime is then $N/2$ executions (for an explanation, see e.g. Supplementary Note 1 of \cite{inesta_optimal_2023}). Recalling that, in our model, $M$ executions of the single-click protocol are batched into a single unit of time, the decoherence rate of the link is then $\Gamma = 2M/N$. 
The batch size in current experiments is $M\approx 500\text{--}1000$ \cite{pompili_experimental_2022, stolk2024metropolitan}, but since this is not a hardware parameter we regard it as freely adjustable.
See Table \ref{table:params_expts} for the specific choices of parameters for both the near-term and far-term regimes.
The true timescale of an attempt according to the parameter regimes in Table \ref{table:params_expts} is on the order of ms \cite{stolk2024metropolitan,pompili_experimental_2022}. We refer to Appendix \ref{app:timing} for more details.
All reported results are in abstract time units. 

For some entanglement generation schemes, adjusting the generation parameters may incur a further time delay before each entanglement generation attempt. In such cases, it is important to take this into account when comparing adaptive and constant-action policies. However, in Appendix \ref{app:timing} we show that this delay is negligible for the batched single-click protocol.

We note that the parameter choices put a restriction on the number of links $n$ it is possible to have simultaneously in memory because necessarily $n\leq t_{\max}$, where $t_{\max}$ is the maximum TTL from (\ref{eqn:t_max_expts}). Then, a reduced $\Gamma$ will increase $t_{\max}$, thereby allowing for more links to be present in memory. We note that, if $n$ is increased, correspondingly each node must also contain at least $n$ high-quality memories, which is experimentally challenging.

\begin{table}[t]
	\caption{Parameters for near-term and far-term regimes}
    \label{table:params_expts}
	\centering
	\begin{tabular}{p{0.3\linewidth} | p{0.25\linewidth} | p{0.15\linewidth}}
        \toprule
        Parameter & Near-term & Far-term  \\
        \midrule
		$N$ (memory lifetime in number of single-click executions) & $5263.15 \newline \approx 5300$ \cite{hermans2022qubit} & 20000 \\
        \midrule
		$p_{\mathrm{det}}$ (photon detection probability) & $5\times 10^{-4} \newline \approx 4.4 \times 10^{-4}$ \cite{hermans_entangling_2023} & $5\times 10^{-4}$\\
        \midrule
        $M$ (batch size) &  500  & 1000 \\
        \midrule
		$\lambda = 1/(2p_{\mathrm{det}}M)$ (batched single-click trade-off parameter (\ref{eq:trade-off})) & $2$ & $1$\\
        \midrule
            $\Gamma = 2M/N$ (decoherence rate, (\ref{eqn:decoherence})) &  $0.19$  & $0.1$\\
        \bottomrule
	\end{tabular}
	\label{tab:milestones}
\end{table}

Depending on the parameter regime, policy performance is computed with one of three different methods: solving the Bellman equations (\ref{eq:bellman}), the formula for completion time from \cite{davies_tools_2024} (for constant-action policies only), or with Monte Carlo simulations. Only in the third case do we need to include error bars, because the first two are analytical formulae. We only use Monte Carlo simulations when the analytical formulae become intractable due to the increase in the size of the state set. In the parameter regimes explored, we only resort to Monte Carlo simulations for the far-term regime with $n\geq 9$ for the heuristic policy $\pi_{\mathrm{h}}$. In these cases, the simulations are calculated as the mean of $10^4$ runs and error bars are computed as confidence intervals with three standard deviations. Error bars are included in the plots, but are too small to be seen.

We ran all experiments on an Asus ROG G14-GA401QM with an AMD Ryzen 9 5900HS (8-core CPU), 16 GB unified memory, and a 1 TB SSD. The limited memory and CPU resources restricted the size and complexity of our workloads. All code used in the experiments is publicly available \cite{gitlab}.

\subsubsection{Near-term regime}
\label{sec:near_term}
The parameters for this regime are ${\Gamma = 0.19}, ~{\lambda = 2}, ~{F_\mathrm{app} = 1/2}$ (see also Table \ref{table:params_expts}). The action set $\mathcal{A}$ is given by (\ref{eqn:actions_batch_single_click}).
For the near-term regime, from (\ref{eqn:t_max_expts}) we have $t_{\max}=6$, which is the maximum value of $n$.

\begin{figure}[t]
    \centering
    \includegraphics[width=\linewidth]{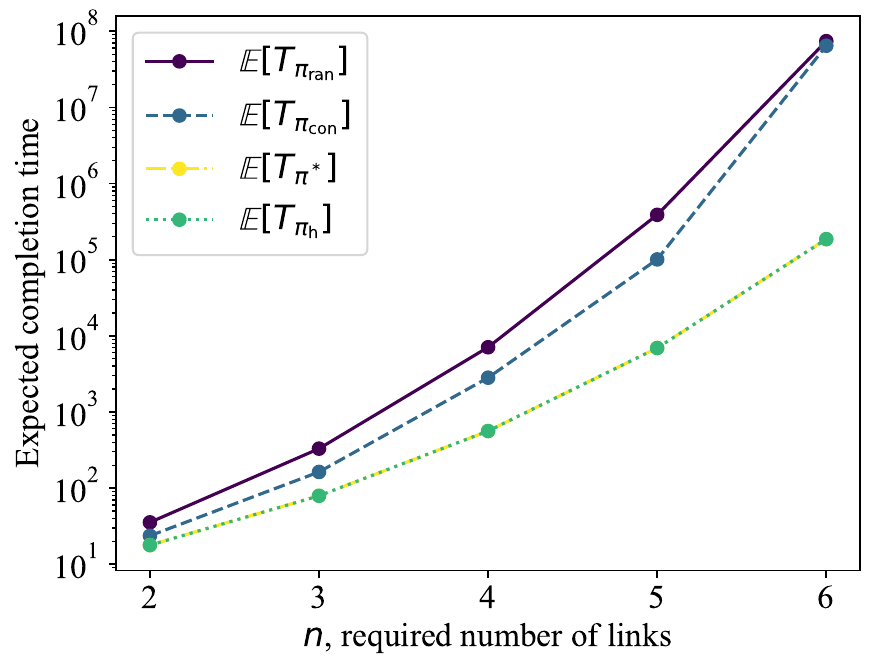}
    \caption{\textbf{Policy performance for the near-term regime.} The parameters for the near-term regime are $\Gamma =0.19$, $\lambda=2$, and $F_{\mathrm{app}}=1/2$ (see Table \ref{table:params_expts} and explanation in the main text). 
    Plotted are the expected completion time for the optimal policy $\pi^*$, the heuristic policy $\pi_\mathrm{h}$, the uniformly random policy $\pi_\textrm{ran}$ and the constant-action policy $\pi_\textrm{con}$. Time units are the number of time steps. There are no error bars as all points were computed either analytically or with policy evaluation.
    }
    \label{fig:small_env_log_performance}
\end{figure}

\begin{figure}[t]
    \centering
    \includegraphics[width=\linewidth]{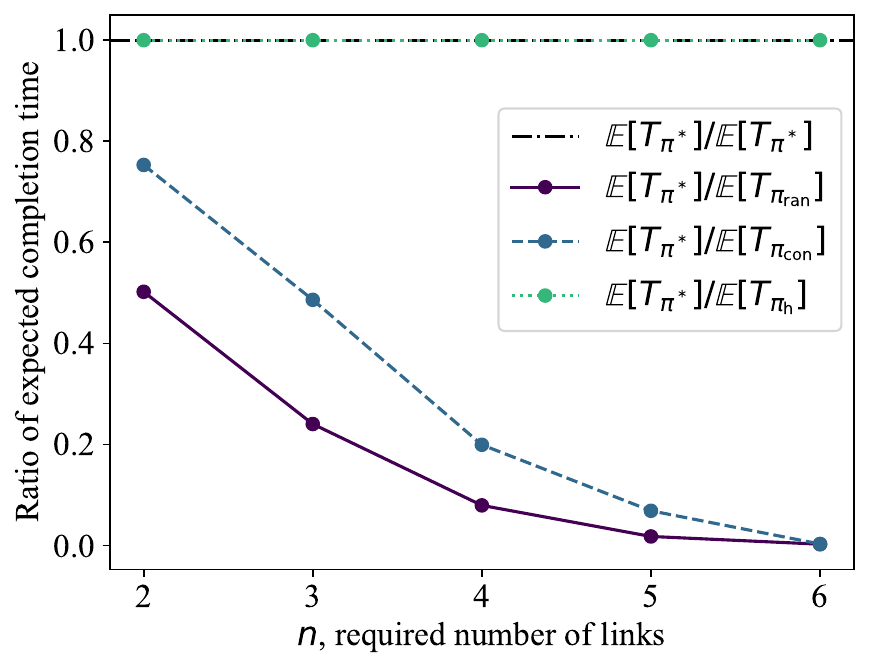}
    \caption{\textbf{Relative policy performance for the near-term regime.}
    Plotted are the ratio of the performance of the optimal policy $\pi^*$ to the performance of the heuristic policy $\pi_\mathrm{h}$ and baselines $\pi_\textrm{ran}$, $\pi_\textrm{con}$. There are no error bars as all points were computed either analytically or with policy evaluation.
    }
    \label{fig:small_env_ratio_performance}
\end{figure}

\autoref{fig:small_env_log_performance} presents the expected completion times for the optimal policy, heuristic policy, and baselines.
In this regime, the performance of all policies is computed up to the maximum $n = t_{\max}=6$.
The heuristic policy and the random-action policy performance are computed exactly by solving the corresponding Bellman equations (\ref{eq:bellman}). We note that, for the random-action policy, the Bellman equations currently written in (\ref{eq:bellman}) require a small generalisation to non-deterministic policies, which can e.g. be found in Chapter 3 of \cite{sutton_reinforcement_markov}.
The optimal policy and its performance are computed with policy iteration (see Section~\ref{sec:policy_iteration}).
The performance of the constant-action policy is computed for all values of $n$ with analytical methods from \cite{davies_tools_2024}.

Remarkably, for all values of $n$ we find that the heuristic policy is optimal,  i.e. $\mathbb{E}[T_{\pi_{\mathrm{h}}}] = \mathbb{E}[T_{\pi^*}]$. This is already expected for $n=2$, because in that regime our heuristic is identical to the analytical solution for the optimal policy as presented in Section \ref{sec:analysis_n=2}. However, for $n>2$, the state set and transitions are more complex and one must use dynamic programming to show that a policy is optimal. In fact, the optimal policy turns out to match exactly with the heuristic policy for all $n$. As multiple optimal policies can exist, this was not necessarily the case.
From the figure, we see that the improvement of the optimal policy (as well as the heuristic policy) over the optimal policy is approximately two orders of magnitude for $n=6$.
To see this improvement more clearly, Figure \ref{fig:small_env_ratio_performance} shows the ratio in performance of the optimal policy with the two baselines and the heuristic policy. As previously discussed, the heuristic policy provides optimal performance for the values of $n$ investigated, meaning that the ratio is one. The advantage in performance increases with $n$, with the optimal (and heuristic) policies providing a performance increase of up to a factor of $\mathbb{E}[T_{\pi_{\mathrm{con}}}]/ \mathbb{E}[T_{\pi^*}] \approx 346$ for the constant-action policy and $\mathbb{E}[T_{\pi_{\mathrm{ran}}}]/ \mathbb{E}[T_{\pi^*}] \approx397$ for the random-action policy.

\begin{figure}[t]
    \centering
    \includegraphics[width=\linewidth, trim=10px 5px 10px 10px]{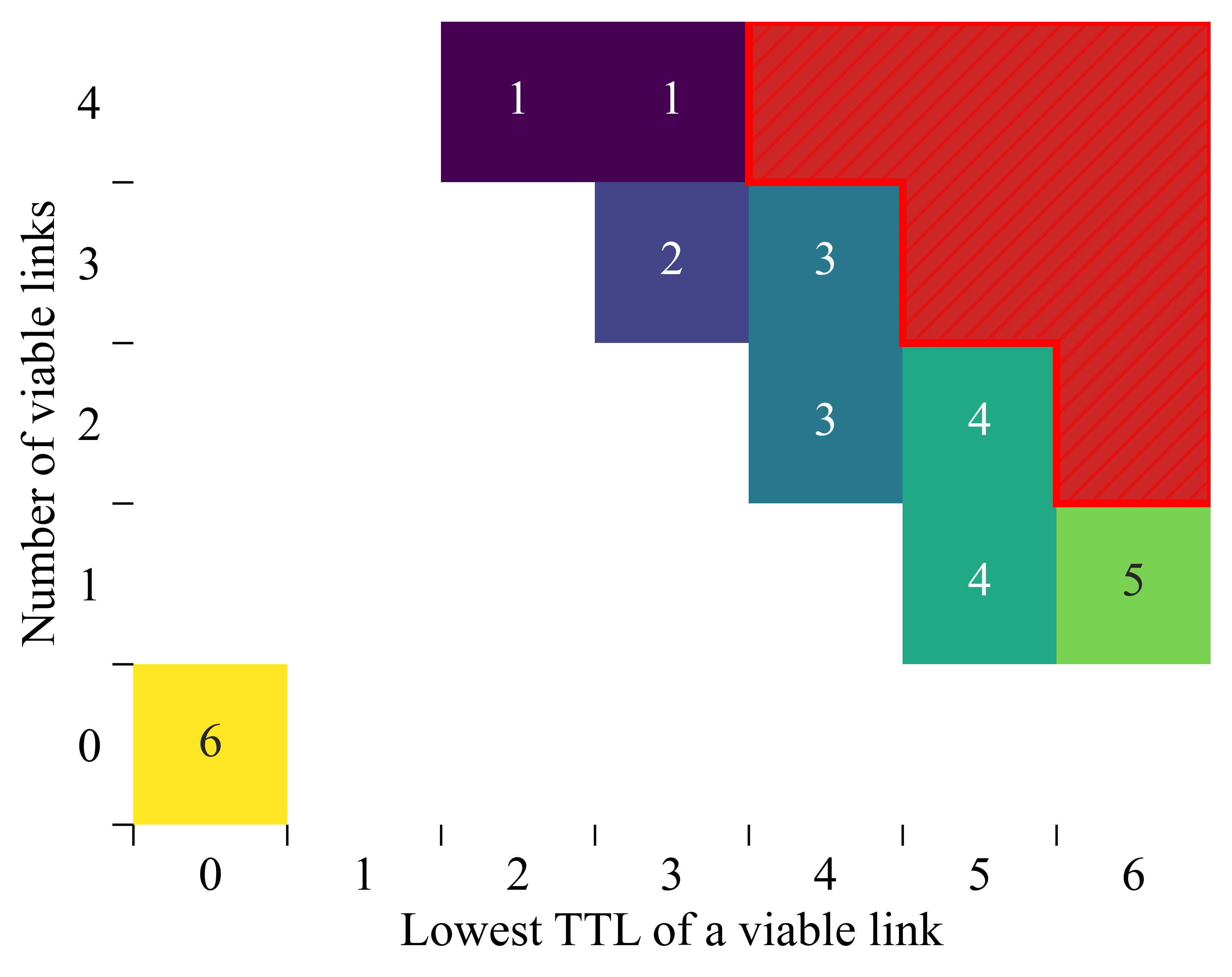}
    \caption{\textbf{Actions chosen by the optimal policy $\pi^*$ for the near-term regime and $n=5$ required links.} The $x$-axis indicates the lowest TTL of a link in a state $s$, $\min \{t: t\in v(s) \} $ from (\ref{eqn:v_definition}). The $y$-axis indicates the number of viable links $N_v(s)$ from (\ref{eqn:N_v_definition}). We note that there can be multiple states that take the same value on both the $y$-axis and $x$-axis. In the heat map, the most commonly-chosen action is shown for all states taking those values. A darker colour (lower number) means that the optimal policy prioritises success probability instead of fidelity. The number in each box is the TTL of the generated link corresponding to the most-commonly chosen action. We see that the optimal policy prioritises the success probability in states with more viable links and states with viable links that will expire more quickly. We note that certain states in the heat map marked in red are inaccessible, such as states with more than one viable link with the lowest TTL being six, since only a single link can be generated at a time. 
     }
    \label{fig:small_env_heatmap}
\end{figure}

In Figure \ref{fig:small_env_heatmap}, the structure of the optimal policy is shown as a heat map for $n=5$. Specifically, it is shown how the action $\pi^*(s)$ depends on the number of viable links in the state $N_v(s)$ and the minimum TTL of a viable link ${\min \{t: t\in v(s) \} }$.
We verify numerically that the optimal policy exactly matches the heuristic policy. From the heat map, we see that the optimal policy usually chooses high-probability (low-fidelity) actions for states containing more viable links and states containing viable links that will expire more quickly. By contrast, when there are zero viable links, the optimal policy increases the fidelity as much as possible at the expense of the success probability, choosing the action that generates a link with the maximum TTL $t_{\max} = 6$. 

We note that certain states in the heat map  are inaccessible, such as when there is more that one viable link with a large TTL. These states have been visually marked for clarity. Such a state is inaccessible due to the assumption of sequential entanglement generation: once a link is generated with maximum TTL, it will necessarily decay as new links are generated. In a system where spatial multiplexing is implemented, allowing for multiple links to be successfully generated per time step, one would have access to such states. This would constitute an interesting extension of our problem.

\subsubsection{Far-term regime}
\label{sec:far_sterm}
The parameters for this regime are ${\Gamma = 0.1},~{\lambda = 1}, ~{F_\mathrm{app} = 1/2}$  (see also Table \ref{table:params_expts}). The action set $\mathcal{A}$ is given by (\ref{eqn:actions_batch_single_click}).
For the far-term regime, from (\ref{eqn:t_max_expts}) we have $t_{\max}=11$, which is the maximum value of the required number of links $n$. In the near-term regime, the maximum number of required links was $n=6$. We thus see that the far-term regime has a larger state set for the same $n$ (see Appendix \ref{sec:statespacesize}).

\begin{figure}[t]
    \centering
    \includegraphics[width=\linewidth]{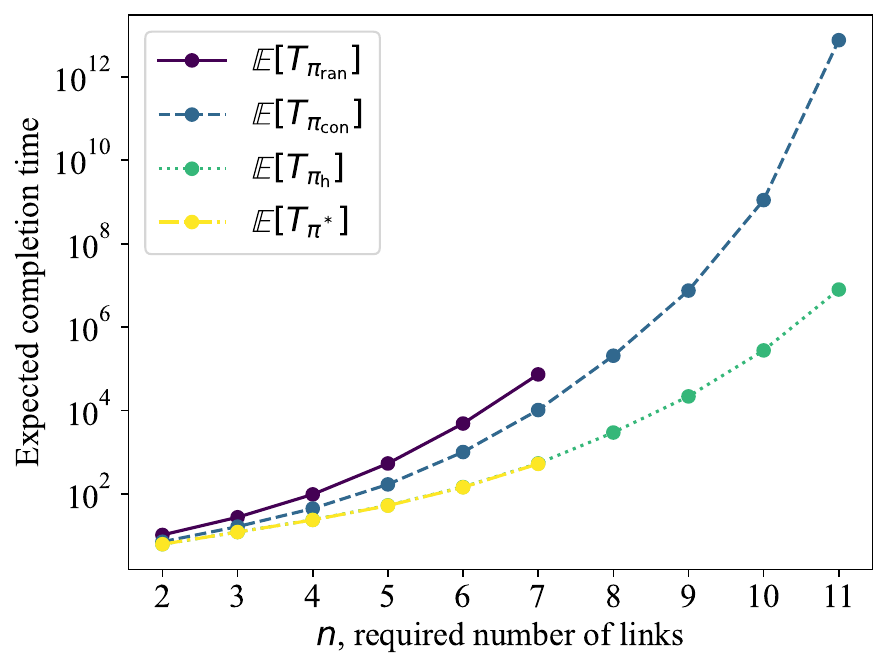}
    \caption{\textbf{Policy performance for the far-term regime.} The parameters for the far-term regime are $\Gamma =0.19$, $\lambda=2$, and $F_{\mathrm{app}}=1/2$ (see Table \ref{table:params_expts} and explanation in the main text). 
    Plotted are the expected completion time for the optimal policy $\pi^*$, the heuristic policy $\pi_\mathrm{h}$, the uniformly random policy $\pi_\textrm{ran}$ and the constant-action policy $\pi_\textrm{con}$. Time units are the number of time steps. Error bars are included for simulated values of $\mathbb{E}[T_{\pi_\textrm{h}}]$ at $n=9,10,11$ with a confidence interval of three standard deviations, but are too small to be visible.
    }
    \label{fig:large_env_log_performance}
\end{figure}

\begin{figure}[t]
    \centering
    \includegraphics[width=\linewidth]{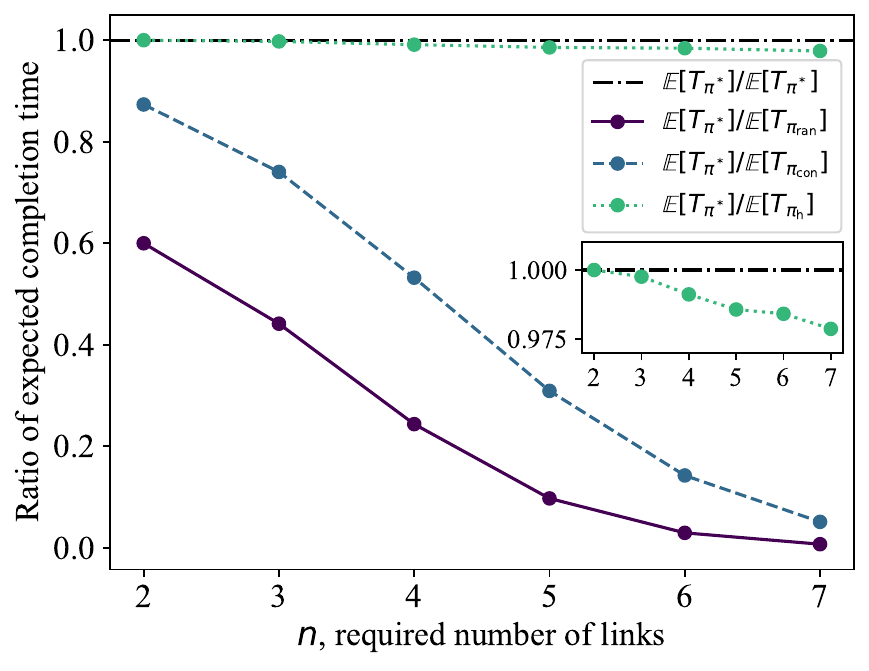}
    \caption{\textbf{Relative policy performance for the far-term regime.}
    Plotted are the ratio of the performance of the optimal policy $\pi^*$ to the performance of the heuristic policy $\pi_\mathrm{h}$ and baselines $\pi_\textrm{ran}$, $\pi_\textrm{con}$. There are no error bars as all points were computed either analytically
or with policy evaluation.
    }
    \label{fig:large_env_ratio_performance}
\end{figure}

Figure \ref{fig:large_env_log_performance} presents the expected completion time for the optimal policy, heuristic and baselines in the far-term regime. Due to the scaling of the state set, we are only able to compute optimal policies for $n\leq 7$. The random-action policy is computed with policy evaluation for $n\leq7$. The performance of the constant-action policy is computed analytically for all $n$ with methods from \cite{davies_tools_2024}.
The performance of the heuristic policy is computed with policy evaluation for $n\leq 8$ and Monte Carlo simulation for $n=9,10,11.$

From Figures \ref{fig:large_env_log_performance} and \ref{fig:large_env_ratio_performance}, we again observe that the heuristic policy maintains a remarkably close performance to the optimal policy for all values of $n$.
As $n$ increases, we again see a greater advantage provided by the optimal policy and heuristic policy over the baselines.
For $n=7$ we see the maximum improvement, where the optimal policy improves on the performance of both the constant-action policy by a factor of $\mathbb{E}[T_{\pi_{\mathrm{con}}}]/ \mathbb{E}[T_{\pi^*}] \approx19$ and random-action policy by a factor of $\mathbb{E}[T_{\pi_{\mathrm{ran}}}]/ \mathbb{E}[T_{\pi^*}] \approx139$. 

We saw in the near-term regime that the heuristic policy is in fact optimal. In the far-term regime, we find that the heuristic is optimal for $n=2$, which again is expected because for $n=2$ the heuristic matches the analytical optimal policy from Section \ref{sec:analysis_n=2}. For $n>2$, we find that the heuristic policy is no longer optimal but still exhibits strong performance. Although the lines for the optimal policy and heuristic policy appear to overlap in Figure \ref{fig:large_env_log_performance}, we see in Figure \ref{fig:large_env_ratio_performance} that for large $n$, the heuristic policy exhibits slightly worse performance than the optimal value. However, the deviation is not significant, and for all $n\leq 7$ the ratio of the two performances satisfies
    \begin{equation}
         \frac{\mathbb{E}[T_{\pi_{\mathrm{h}}}] - \mathbb{E}[T_{\pi^*}]}{\mathbb{E}[T_{\pi^*}]} < 0.03.
    \end{equation}

\begin{figure}[t]
    \centering
    \includegraphics[width=\linewidth,trim=10px 10px 10px 10px]{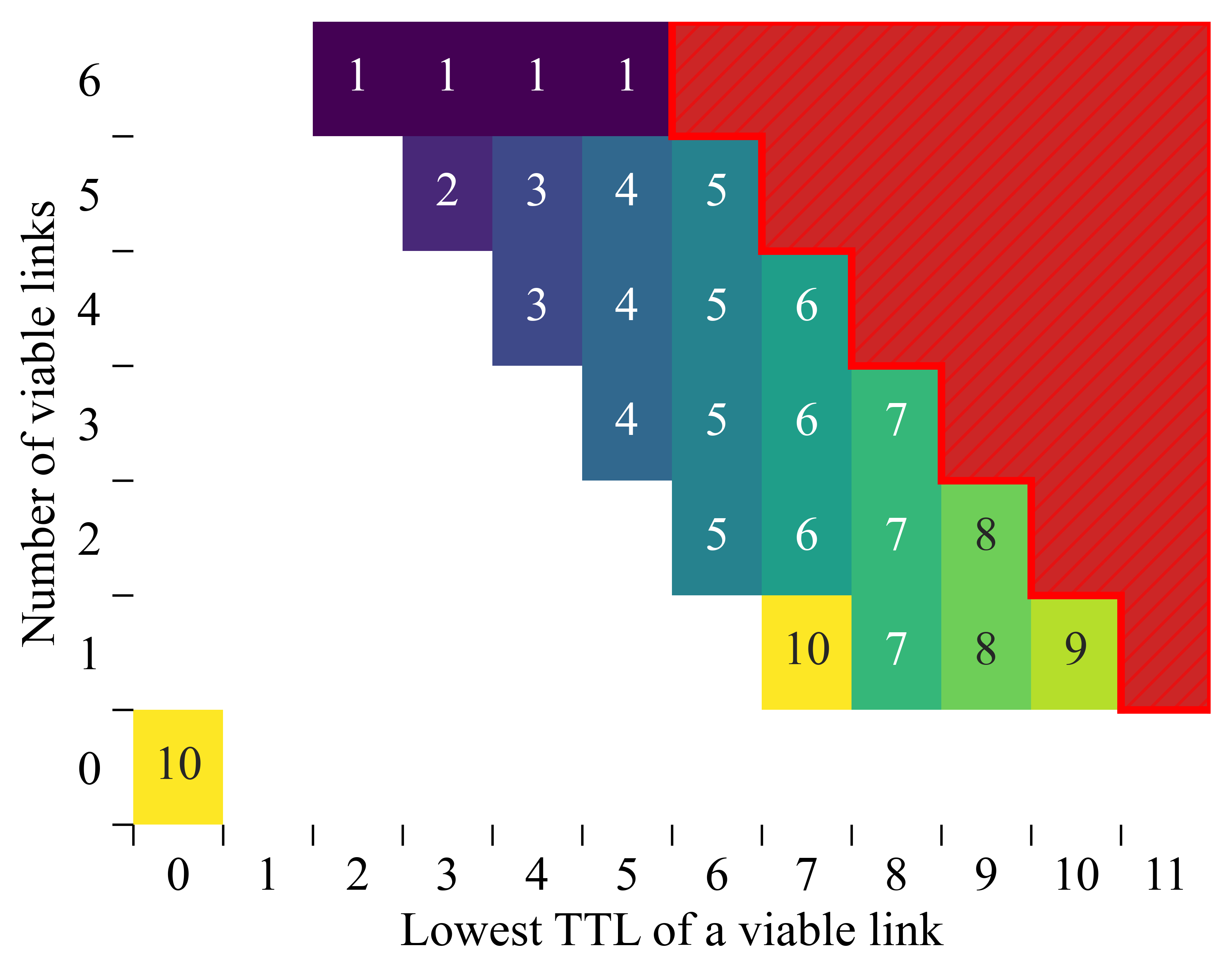}
    \vspace{-1.5em}
    \caption{\textbf{Actions chosen by the optimal policy $\pi^*$ for the far-term regime and $n=7$ required links.} The $x$-axis indicates the lowest TTL of a link in the state $s$, $\min \{t: t\in v(s) \} $ from (\ref{eqn:v_definition}). The $y$-axis indicates the number of viable links $N_v(s)$ from (\ref{eqn:N_v_definition}). We note that there can be multiple states that take the same value on both the $y$-axis and $x$-axis. In the heat map, the most commonly-chosen action is shown for all states taking those values. A darker colour means that the optimal policy prioritises success probability instead of fidelity. The number in each box is the TTL of the generated link corresponding to the most-commonly chosen action. We see that the optimal policy prioritises the success probability in states with more viable links and states with viable links that will expire more quickly. However, there are some outliers, such as states with one viable link with a TTL of seven. We note that certain states in the heat map marked in red are inaccessible. This includes states with more than one viable link with the lowest TTL being $11$, since only a single link can be generated at a time, as well as links with a TTL higher than the largest-TTL action (that generates links with TTL $10$), which are never generated by the policy.}
    \vspace{-1.5em}
    \label{fig:large_env_heatmap}
\end{figure}

Figure \ref{fig:large_env_heatmap} visualises the optimal policy as a heat map for the case where the required number of links is $n=7$. 

We notice that, for states with no viable links, the optimal policy chooses to generate a link with a TTL of ten, even though it is in principle possible to generate a link with a TTL of 11.
We therefore see that the TTL (fidelity) should not necessarily be maximised when generating the first link, because the generation probability might be sacrificed too much.
We already account for this in our heuristic policy, by making the initial action arbitrary in the step (\ref{eqn:heuristic_policy_zero_viable}) and later optimising over this parameter.

We again see the same patterns also seen in Figure \ref{fig:small_env_heatmap}: the optimal policy usually chooses high-probability (low-fidelity) actions for states containing more viable links and states containing viable links that will expire more quickly. This is also captured by the heuristic.
However, in the far-term regime we also see these rules broken in certain circumstances.
For example, consider a state $s$ with one viable link $N_v(s) = 1$ with TTL $t_{N_v(s)}=7$. In the state $s$, we see from Figure \ref{fig:large_env_heatmap} that the optimal policy most commonly chooses the same generation parameters as it would for states with no viable links. 
A potential reason for this is that even though the single viable link may still survive until the remaining links are generated, the probability that it does so is very low. 
Thus, we learn that sometimes it is worth abandoning viable links that have a low TTL and directly start generating new links with a high TTL. 
This is something that the heuristic does not account for, and we identify this as the reason why the heuristic is sub-optimal. 
Extending our heuristic to account for this is a potential avenue for improvement. 

\subsubsection{Effect of changing $F_\mathrm{app}$}
\label{sec:changeFapp}
\begin{figure}[t]
    \centering
    \includegraphics[width=\linewidth]{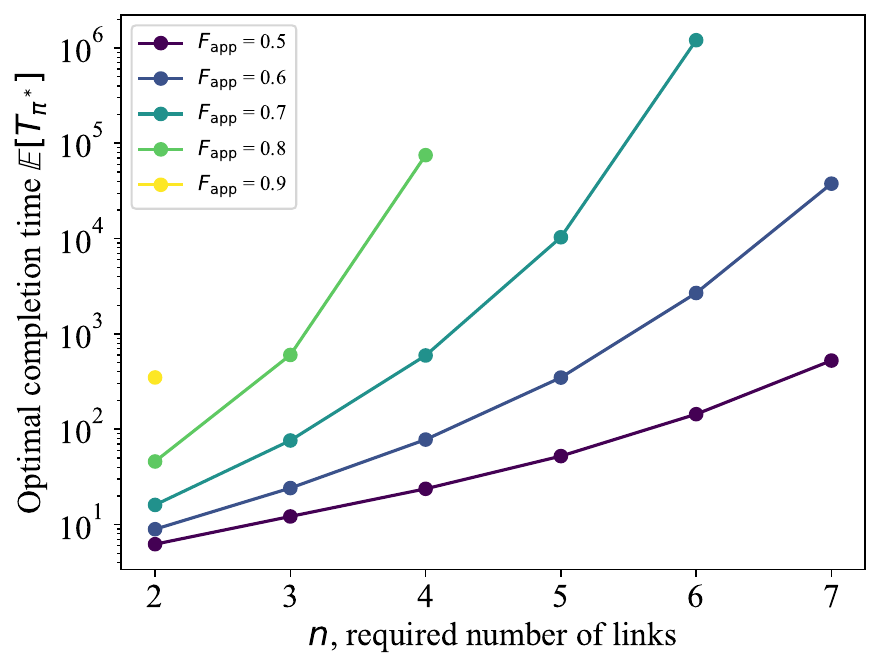}
    \caption{\textbf{Optimal policy performance for the far-term regime with a varying $F_{\mathrm{app}}$.} The parameters for the near-term regime are $\Gamma =0.1$, $\lambda=1$ (see Table \ref{table:params_expts} and explanation in the main text). 
    We plot the expected completion time for the optimal policy $\pi^*$. 
    }
    \label{fig:policy_performance_varying_F_ap}
\end{figure}

We now analyse the effect of changing $F_\mathrm{app}$, the minimum required fidelity for each link, on the performance of the optimal policy. We again compare with the heuristic policy and the baselines. We carry out the analysis for the far-term regime only. This is because, in the near-term regime, there is only a small feasible range of $F_\mathrm{app}$ before generating an entanglement packet with $n\geq 2$ links is no longer possible. We recall that the maximum size of an entanglement packet is given by the maximum TTL $t_{\max}$ via (\ref{eqn:TTL}) and (\ref{eqn:t_max_expts}).
We therefore compare policy performance in the far-term regime, where entanglement packets with $n\geq 2$ can be generated for a wider range of $F_\mathrm{app}$ (see Figure \ref{fig:policy_performance_varying_F_ap}, where $0.5 \leq F_\mathrm{app}\leq 0.9$).

In Figure \ref{fig:policy_performance_varying_F_ap}, it is shown how the performance of the optimal policy $\mathbb{E}[T_{\pi^*}]$ varies with the number of required links $n$ in the entanglement packet, for varying $F_\mathrm{app}$. For $F_\mathrm{app}=0.9$, it is only possible to generate entanglement packets of maximum size $n=2$.
For each value of $F_\mathrm{app}$, we again see that, as $n$ increases, the optimal completion time necessarily also increases. As $n$ approaches $t_{\max}$, this increase becomes exponential, with the optimal completion times for $F_\mathrm{app}\geq 0.6$ experiencing an order-of-magnitude increase or more. For lower $F_\mathrm{app}$, such as $F_\mathrm{app}=0.5$, we see that the price paid for increasing $n$ is less severe. A reason for this could be that, as the fidelity requirement becomes more stringent, one requires multiple successes in close succession within a window of time bounded above by $t_{\max}$. In \cite{davies_tools_2024} it was shown that, for a constant-action policy with parameters $(p,F)$, the expected completion time scales as $\mathbb{E}[T_{\pi_{\mathrm{con}}}]\sim 1/p^n$ as $p$ approaches zero. In particular, in a hard-to-generate regime with $1/p\gg t_{\max}$, the expected completion time of the constant-action policy appears to increase exponentially with $n$. We conjecture that one may therefore anticipate an exponential increase to also occur with the optimal policy, since one still requires several links to be generated almost successively within a short period of time.

\begin{figure}[t]
    \centering
    \includegraphics[width=\linewidth]{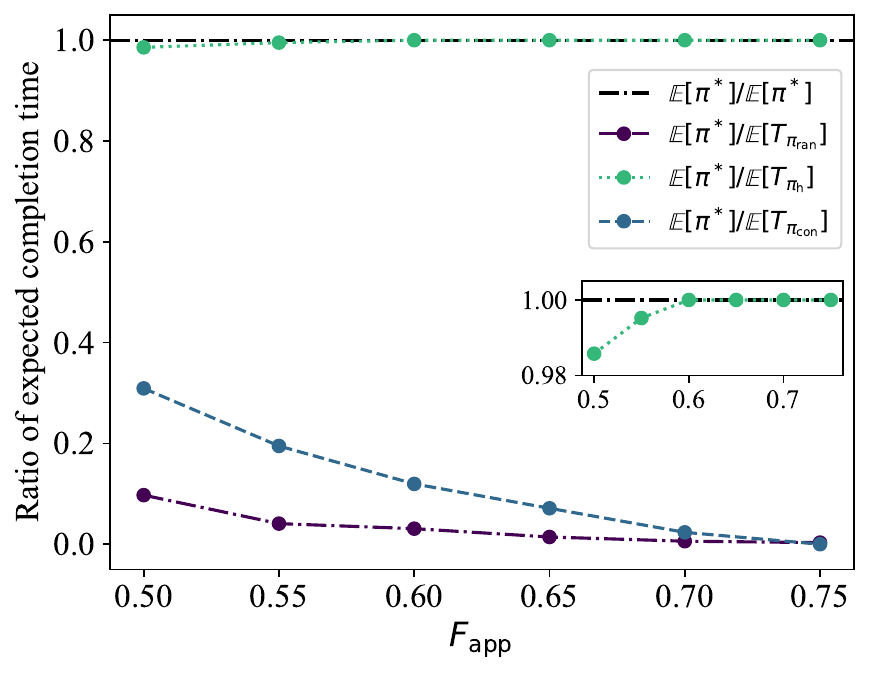}
    \caption{\textbf{Relative policy performance for the far-term regime with a varying $F_{\mathrm{app}}$ and $n=5$ required links.} The parameters for the far-term regime are $\Gamma =0.1$, $\lambda=1$ (see Table \ref{table:params_expts} and explanation in the main text). 
    Plotted are the expected completion time for the optimal policy $\pi^*$, the heuristic policy $\pi_\mathrm{h}$, the uniformly random policy $\pi_\textrm{ran}$ and the constant-action policy $\pi_\textrm{con}$. There are no error bars as all points were computed either analytically
or with policy evaluation.
    }
    \label{fig:ratio_performance_varying_F_app}
\end{figure}

In Figure \ref{fig:ratio_performance_varying_F_app}, we plot the ratio of the optimal expected completion time to the heuristic policies and two baselines, for the far-term regime, $0.5 \leq F_\mathrm{app}\leq 0.75$ and $n=5$. Here, we have chosen $n=5$ so the expected completion time can be analysed for a reasonably large range of $F_\mathrm{app}$. As we saw in Figure \ref{fig:policy_performance_varying_F_ap}, fixing a larger size of entanglement packet ($n=6$ or $n=7$) limits the valid range of $F_\mathrm{app}$.  From Figure \ref{fig:ratio_performance_varying_F_app}, we see that the advantage of the optimal policy over the baselines grows significantly as $F_\mathrm{app}$ increases. We draw a similar conclusion as we did in the previous sections (the near-term and far-term regime for $F_\mathrm{app}=0.5$): in a harder-to-generate regime that has a more stringent fidelity requirement, adaptive policies show a greater advantage than non-adaptive policies. We notice from the same figure that the heuristic policy is near-optimal for $F_\mathrm{app}\leq 0.55$ and optimal for $F_\mathrm{app} >0.55$.

\begin{figure}[t]
    \centering
    \includegraphics[width=\linewidth]{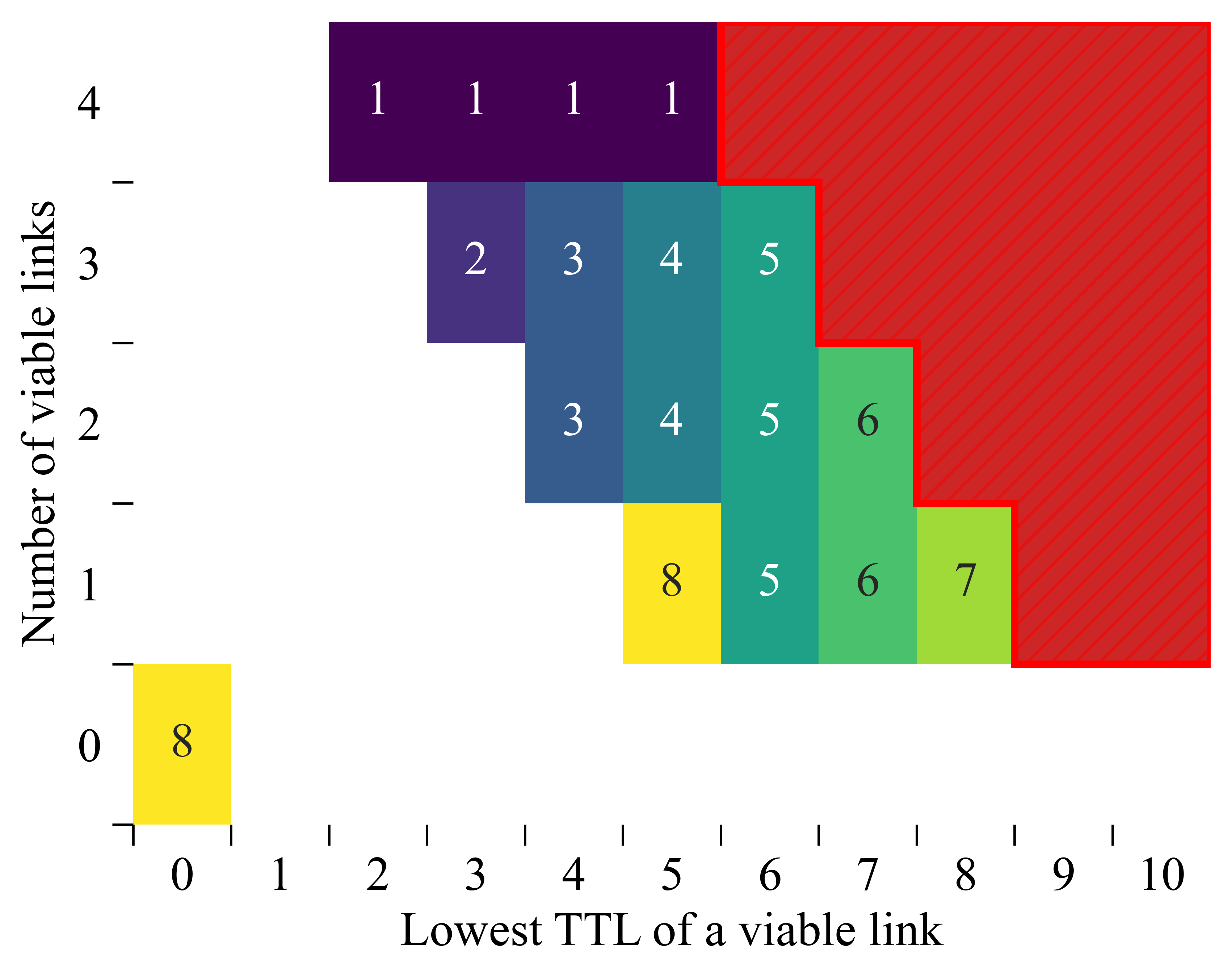}
    \caption{\textbf{Actions chosen by the optimal policy $\pi^*$ for the far-term regime with $F_{\mathrm{app}}=0.55$ and $n=5$ required links.} The $x$-axis indicates the lowest TTL of a link in the state $s$, $\min \{t: t\in v(s) \} $ from (\ref{eqn:v_definition}). The $y$-axis indicates the number of viable links $N_v(s)$ from (\ref{eqn:N_v_definition}). We note that there can be multiple states that take the same value on both the $y$-axis and $x$-axis. In the heat map, the most commonly-chosen action is shown for all states taking those values. A darker colour means that the optimal policy prioritises success probability instead of fidelity. The number in each box is the TTL of the generated link corresponding to the most-commonly chosen action.
    }
    \label{fig:varying_F_app_heatmap}
\end{figure}

The heat map in Figure \ref{fig:varying_F_app_heatmap} displays the structure of the optimal policy for the far-term regime with $F_{\mathrm{app}}=0.55$ and $n=5$ required links. We see a similar structure to the optimal policy as depicted in Figure \ref{fig:large_env_heatmap} (far-term regime, $F_\mathrm{app}=0.5$, $n=7$): when there are few viable links with a low TTL, the optimal policy sometimes chooses to abandon them, by instead choosing generation parameters with a small success probability and large TTL. In Figure \ref{fig:varying_F_app_heatmap}, we see this behaviour occurring for states with a single viable link that has a TTL of five. We conclude that the abandoning behaviour, which was identified in the previous section as the reason why the optimal policy performs slightly better than the heuristic, remains a structural feature of the optimal policy for $F_\mathrm{app}=0.55$, and this again explains why the heuristic is not optimal in this parameter regime. As discussed previously, for larger value of $F_\mathrm{app}$ in the parameter regimes explored, the heuristic is again optimal.  An explanation for this could be the following: that there are fewer actions with a high TTL as $F_\mathrm{app} $ increases, meaning that is less worthwhile to sacrifice older links to try to generate a new high-fidelity batch. 

Since multiple optimal policies may exist, we cannot rule out the existence of an equally optimal policy without the abandoning behaviour. The observation of the abandoning effect across two different $F_\mathrm{app}$ values could suggest that the abandoning behaviour is a structural feature of an optimal policy. However, we emphasise that this is a tentative conclusion only drawn from the parameter regimes considered in this work, and to investigate its robustness one would have to find optimal policies in different parameter regimes, and/or find more examples of optimal policies in the same parameter regime.


\section{Conclusion and future work}
\label{sec:discussion}
We have considered a scenario where entanglement generation attempts are sequential in time and the generation parameters may be chosen in each time step.
By formulating the problem as an MDP, we have found policies that minimise the expected time to generate multiple simultaneously-existing links. In the parameter regimes explored, we observed that our optimal policies provide a significant improvement over the constant-action and random-action baselines.
We have also found a heuristic method to compute policies that exhibit either optimal or close-to-optimal performance in all parameter regimes explored. The heuristic method is more efficient to compute than finding optimal policies with dynamic programming, and we therefore expect it to be useful in situations where the optimal policy cannot be computed due to the scaling of the state set (e.g. when the number of required links $n$ is large). However, we note that this is intuition gained from exploring only the parameter regimes considered in this work, and the conclusion may differ when considering different examples for the $(p,F)$ trade-off or values of the decoherence rate $\Gamma$ and packet size $n$.
Nevertheless, our work highlights that adaptive protocols leveraging the rate-fidelity trade-off inherent to the entanglement generation scheme can be extremely helpful in improving quantum network performance. Moreover, in certain important entanglement generation schemes, the rate-fidelity trade-off is easily tunable (for example, by varying the bright-state parameter or mean photon number \cite{hermans_entangling_2023,knaut2024entanglement}), and so our adaptive protocols are readily implementable. For parameter regimes in which finding the optimal policy is not feasible with dynamic programming, a fruitful direction of research may be deep reinforcement learning, which finds approximate solutions. 

Our model does not assume a specific entanglement generation protocol. In our results, we have used the example of a batched single-click protocol. It would also be interesting to study optimal policies for different entanglement generation schemes. A different entanglement generation scheme may result in a different action set, and therefore optimal policies with distinct properties to the batched single-click case.

Given the substantial speed-up we find with adaptive entanglement generation in this work, one may also expect an improvement for other important quantum network protocols, aside from the generation of entanglement packets. One example is that of a quantum repeater chain, where multiple links must be generated simultaneously, not between two nodes, but between several segments of a linear chain of repeaters in order to be transformed into end-to-end entanglement through entanglement swapping \cite{dur1999quantum,inesta_optimal_2023}. Another example is that of an entanglement switch, where multiple links are generated between a central node and several users, in order to be later transformed into multipartite entanglement with a GHZ measurement \cite{nain2020analysis,avis_requirements_2023,van2024utilizing}. It would also be interesting to compare or combine adaptive policies with other methods to improve entanglement distribution. An interesting example is spatial multiplexing, which would allow for multiple successes to occur within one time step. In order to study this within the MDP framework, one would therefore need to update the state set, transition function and action set to account for this, and potentially need to account for adapting several entanglement generation parameters at every time step.
Given the advantage that we saw was gained through adaptive entanglement generation, we envisage that this might influence future network architecture design decisions. For example, it may be advantageous to employ adaptive entanglement generation schemes (such as the single-click scheme, as considered in detail in this work) over non-adaptive schemes (such as the double-click scheme \cite{barrett_efficient_2005}). 
However, in order to confirm this intuition, it is important to study policies for adaptive entanglement generation in different parameter regimes and for a wider variety of quantum network subroutines.

\section*{Acknowledgements}
We thank Eric Bersin for helpful discussions, and Álvaro G. Iñesta for critical feedback on the manuscript.
This work is supported in part by QuTech NWO funding 2020-2024 Part I “Fundamental Research”, Project Number 601.QT.001-1, financed by the Dutch Research Council (NWO). We further acknowledge support from NWO QSC grant BGR2 17.269 and NSF grant \#2522101.
BD acknowledges financial support from the Ammodo foundation.

\appendices

\section{Flow of events within one time step}
\label{app:timing}
Here, we outline the events assumed to occur in each time step. These events determine the transitions of the Markov decision process in Section \ref{sec:mdp_construction}. In a single time step, the following six events occur consecutively.
\begin{enumerate}[label=\roman*.]
    \item Based on the state of the system, which is defined by the number of links currently in memory and their corresponding fidelities, the system chooses the generation parameters $(p,F)$ to use for entanglement generation. This takes time $\Delta t_{\mathrm{par}}$.
    \item The system adjusts the necessary hardware and/or protocol parameters, which fixes the generation parameters to be equal to $(p,F)$. This step takes time $\Delta t_{\mathrm{adj}}$.
    \item Entanglement generation is attempted with parameters $(p,F)\in \mathcal{A}$. This step takes time $\Delta t_{\mathrm{gen}}$. 
    \item All links experience decoherence over the time $\Delta t_{\mathrm{par}} + \Delta t_{\mathrm{adj}} + \Delta t_{\mathrm{gen}}$ according to (\ref{eqn:decoherence}).
    \item With probability $p$, a fresh link with fidelity $F$ is generated.
    \item Links that have decohered below $F_{\mathrm{app}}$ are discarded, taking time $\Delta t_{\mathrm{dis}}$.  
\end{enumerate}
The duration of a single time unit is then $\Delta t = \Delta t_{\mathrm{par}} + \Delta t_{\mathrm{adj}} + \Delta t_{\mathrm{gen}} + \Delta t_{\mathrm{dis}}$. We note that here, it is not assumed that decoherence occurs during the discarding time. This is valid for central quantum communication platforms such as Nitrogen-vacancy (NV) centres, where the dominant source of decay occurs due to entanglement generation attempts causing decoherence of states stored in memory \cite{reiserer2016robust}. However, decoherence during the discarding time may be taken account in the model by assuming an increased decoherence rate $\Gamma$ and a choice of generation schemes with reduced-fidelity generation pairs $(p,F_{\mathrm{red}})$, where $F_{\mathrm{red}}$ is the fidelity of the newly-generated entangled state after decohering for time $\Delta t_{\mathrm{dis}}$. 

We now analyse the dominant sources of time delay within a given time step. In step (i), the generation parameters $(p,F)$ will be chosen according to a pre-computed policy, resulting in no extra computational overhead for making decisions on the fly. Let us assume that each node has full access to the noise parameters of that quantum register (i.e. the decoherence map (\ref{eqn:decoherence})). Since entanglement generation attempts are heralded, meaning that each node is aware when entangled links are successfully generated, the state of the system is fully available to each node at the end of each time step. Therefore, we may assume that step (i) takes negligible time, such that $\Delta t_{\mathrm{par}} = 0$. 
We now consider the time taken to carry out step (ii). 
Let us take as a primary example a batched single-click protocol, which is considered explicitly for our experiments in Section \ref{sec:results}. 
There, the operation required to carry out step (ii) is that each of the two nodes locally adjusts its bright-state parameter, which corresponds to a local rotation on the spin of the communication qubit (i.e. the qubit used to generate remote entanglement).
In the setting of a quantum network, local operations are typically orders of magnitude faster than network operations such as entanglement distribution. 
For example, in refs. \cite{pompili_experimental_2022,stolk2024metropolitan}, entanglement distribution is carried out with a batched single-click protocol. There, NV centres in diamond are used as the platform to perform quantum communication.
In refs. \cite{pompili_experimental_2022,stolk2024metropolitan}, the time taken for a single entanglement generation attempt is on the order of 1-10 $\mu$s.
Contained within this time is the communication qubit initialisation, the duration of which is the execution time of a local gate on an NV electron spin, which is on the order of 10-100 ns \cite{dahlberg2019link,avis_requirements_2023,vandervecht2025qoala}. 
Furthermore, in the network-stack demonstration of \cite{pompili_experimental_2022}, entanglement generation is performed with batches of 1000 individual single-click attempts in each time step, which has a duration on the order of ms.
In \cite{pompili_experimental_2022}, the bright-state parameter is reconfigured repeatedly in real time. 
No associated overhead appears in the reported latency breakdown (Fig. 4b of \cite{pompili_experimental_2022}), which is dominated by entanglement generation and the charge-and-resonance check, and so we conclude that the adjustment of the bright-state parameter causes minimal overhead and therefore assume that $\Delta t_{\mathrm{adj}}=0$ in our analysis. 
Finally, the time $\Delta t_{\mathrm{dis}}$ to discard an entangled pair constitutes a reinitialisation of the corresponding pair of memory qubits and is fully local. 
For networks of NV centres, the corresponding memory qubit is the Carbon-13 spin, which has a reinitialisation time on the order of 100 $\mu$s \cite{dahlberg2019link}.
Assuming that this delay is included in every attempt, regardless of whether entanglement is being discarded or not, every time step has the same length $\Delta t \approx \Delta t_\mathrm{gen} + \Delta t_\mathrm{dis}$. For the important example of the batched single-click protocol with NV centres, given the above explanation, $\Delta t$ is on the order of ms.
We therefore see that the time step length $\Delta t$ is approximately equal for both adaptive and non-adaptive policies, since $\Delta t_\mathrm{gen}$ and $\Delta t_\mathrm{dis}$ are the same.

When applying the analysis to an entanglement generation scheme other than batched single-click, which may have a non-negligible $\Delta t_{\mathrm{adj}}$, one may have to account for any added latency in the comparison of adaptive versus constant policies. Let us take as an example a scenario where adapting the generation parameters requires long-distance communication, such as when we vary a parameter forming part of an entanglement generation protocol on a repeater chain. The parameter, for example, could be a cut-off time that is applied on every intermediate node \cite{inesta_optimal_2023}. Then, the time $\Delta t_{\mathrm{adj}}$ may include the classical communication of the new generation parameters $(p,F)$ to the component nodes of the chain, thereby taking up non-negligible time. Since all adjustments happen within a single time step, our MDP as defined in Section \ref{sec:mdp_construction} is unchanged, but one must account for the extra latency due to adaptivity when comparing with constant-action policies. In such a scenario, an adaptive policy $\pi_{\mathrm{ad}}$ outperforms a constant-action policy $\pi_{\mathrm{con}}$ if $$\Delta t_{\mathrm{con}} \mathbb{E}[T_{\pi_\mathrm{con}}] > (\Delta t_{\mathrm{con}} + \Delta t_{\mathrm{adj}})\mathbb{E}[T_{\mathrm{\pi_{ad}}}],$$
where $\mathbb{E}[T_{\pi}]$ is the expected completion time of policy $\pi$, $\Delta t_{\mathrm{con}}$ is the time step length for constant-action policies, and $\Delta t_{\mathrm{con}} + \Delta t_{\mathrm{adj}}$ is the time step length for adaptive policies.


\section{State set size}
\label{sec:statespacesize}
Here, we quantify exactly the size of the state set, $|\mathcal{S}|$. We recall from Section \ref{subsec:state_space} that $\mathcal{S} = \bigcup_{m=1}^{n-1} \mathcal{S}_m \cup \{\varnothing\}$, where 
\begin{equation}
    \mathcal{S}_m = \big\{\left\{t_1,\dots, t_m \right\} : t_i \in [t_{\mathrm{max}}] , t_1 \geq \dots \geq t_m \big\}.
\end{equation}
We firstly quantify $|\mathcal{S}_m|$. We note that each state in $\mathcal{S}_m$ corresponds to a unique outcome when choosing an element of $[t_{\mathrm{max}}]$ exactly $m$ times, with replacement but without different permutations. Therefore, we have

\begin{equation}
    |\mathcal{S}_m| = \multiset{t_\textrm{max}}{m},
    \label{eqn:S_m_size}
\end{equation}
where \multiset{n}{k} denotes the $k$-combination with repetitions. Then,
\begin{align}
    |\mathcal{S}| &= |\{\varnothing\}| + \sum_{m=1}^{n-1} |\mathcal{S}_m| \\ &= \sum_{m=0}^{n-1} \multiset{t_{\mathrm{max}}}{m} \\ &= \sum_{m=0}^{n-1} {t_\textrm{max} + m - 1 \choose m}, \label{eqn:state_space_sum_choose}
\end{align}
where in the first step we have used the fact that $\mathcal{S}_i$ and $\mathcal{S}_j$ are mutually exclusive sets for $i\neq j$, in the second step we have used (\ref{eqn:S_m_size}), and in the last step we have used the formula
\begin{equation}
    \multiset{n}{k} = {n+k-1 \choose k}.
    \label{eqn:k_comb_with_rep}
\end{equation}
We can further simplify (\ref{eqn:state_space_sum_choose}) using the hockey-stick identity \cite{jones_hockeystick}, 
\begin{equation}
    \label{eq:hockeystick}
    \sum_{j=0}^{w-r} \binom{j+r}{r} = \binom{w + 1}{w - r}
\end{equation}
for $w\geq r$. Then,
\begin{align}
    |\mathcal{S} |&= \sum_{m=0}^{n-1} \binom{t_\textrm{max}-1+m}{m} \\
    &= \sum_{m=0}^{n-1} \binom{t_\textrm{max}-1+m}{t_\textrm{max}-1} \\
    &= \binom{t_\textrm{max}+n-1}{n-1}.
\end{align}
We can then bound the state set size as follows,
\begin{align}
    {t_\textrm{max}+n-1 \choose n-1} &= \prod_{i=0}^{n-2}\frac{t_\textrm{max}+n-1 - i}{n-1-i} \\ &\geq\prod_{i=0}^{n-2} \frac{t_\textrm{max}+(n-1)}{n-1} \label{eq:state_space_lb_step} \\ &= \left(1+\frac{t_\textrm{max}}{n-1}\right)^{n-1}.
    \label{eqn:state_space_lb}
\end{align}
To obtain (\ref{eq:state_space_lb_step}), we have used the fact that 
\begin{equation}
    y = \frac{A-x}{B-x}
\end{equation}
is an increasing function of $x$, for $A>B$. 

By definition of our problem, we have $t_\textrm{max} \geq n-1$. Then, by (\ref{eqn:state_space_lb}) we see that $|\mathcal{S}|\geq 2^{n-1}$, i.e. the size of the state set scales exponentially with $n$ (within the feasible region $n\leq t_\textrm{max}$).

We now quantify the reduction in the size of the state set that may be obtained by identifying states only with their viable links. Recall the function $N_v$ that counts the number of viable links, defined in (\ref{eqn:N_v_definition}). We also recall the function $v$, such that $v(s)$ only contains the viable links of $s$, defined in (\ref{eqn:v_definition}). We define the reduced state set $\mathcal{S}_\mathrm{red} \subset \mathcal{S}$ by identifying $s \equiv v(s)$ for all $s\in \mathcal{S}$. Then, 
\begin{equation}
    \mathcal{S}_\mathrm{red} = \{s\in  \mathcal{S} : v(s) = s\}.
\end{equation}
We now compute $|\mathcal{S}_\mathrm{red}|$. For $m = 0,\dots,n-1$, we define 
\begin{equation}
    \mathcal{S}_{m,\mathrm{red}}  \coloneqq \{s\in \mathcal{S}_{\mathrm{red}}: N_v(s) = m\} 
    \label{eqn:S_m,red}
\end{equation}
where we note that for $s\in \mathcal{S}_{\mathrm{red}}$, we simply have $|s| = N_v(s)$, or equivalently all links are viable. 
The set $\mathcal{S}_{m,\mathrm{red}}$ is interpreted as the set of states containing exactly $m$ viable links.
We note that $\mathcal{S}_{0,\mathrm{red}} = \{\varnothing\}$. We also note that $\mathcal{S}_{m,\mathrm{red}} \subset  \mathcal{S}_m$.
We then have 
\begin{align}
    \left|\mathcal{S}_\mathrm{red}\right| &= \left| \bigcup_{m=0}^{n-1}\mathcal{S}_{m,\mathrm{red}}  \right| \\ &= 1 + \sum_{m=1}^{n-1} \left|\mathcal{S}_{m,\mathrm{red}}\right|,
    \label{eqn:reduced_state_space_sum}
\end{align}
where in the second step we have used the fact that $\mathcal{S}_{i,\mathrm{red}}$ and $\mathcal{S}_{j,\mathrm{red}}$ are mutually exclusive for $i\neq j$. We now compute $|\mathcal{S}_{m,\mathrm{red}}|$. Recalling from Section \ref{sec:heuristic} that the state $s = \{t_1,\dots, t_m \}$ contains $m$ viable links when $t_m > n-m$, it follows that  
\begin{equation}
    \mathcal{S}_{m,\mathrm{red}} \\ = \{\{t_1, \dots, t_m \} : t_i \in [t_{\mathrm{max}}]\setminus [n-m], t_1 \geq \dots \geq t_m \}.
\end{equation}
Then, we deduce $|\mathcal{S}_{m,\mathrm{red}}|$ by the same counting argument as was used above for $|\mathcal{S}_m|$, to find 
\begin{equation}
|\mathcal{S}_{m,\mathrm{red}}| = \multiset{t_{\mathrm{max}} + m - n}{m}.
\end{equation}
Combining with (\ref{eqn:reduced_state_space_sum}), we then have 
\begin{align}
    |\mathcal{S}_\mathrm{red}| &= 1 + \sum_{m=1}^{n-1} \multiset{t_{\mathrm{max}} + m - n }{m}
   \\ &= 1 + \sum_{m=1}^{n-1} {t_\textrm{max}+ 2m-n -1 \choose m},
\end{align}
where we again made use of (\ref{eqn:k_comb_with_rep}).

\section{Trade-off relation for batched single-click scheme}
\label{sec:tradeoff}
Here, we derive a trade-off relation between $p$ and $F$ for batched attempts of the single-click entanglement generation protocol \cite{cabrillo1999creation}.
In the limit of high photon losses ($\eta \ll 1$), if the success probability of a single execution of the single-click protocol is $p_{\mathrm{succ}}$, the fidelity of a generated link is 
\begin{equation}
    F = 1 - \frac{p_{\mathrm{succ}}}{2p_{\mathrm{det}}},
    \label{eqn:fidelity_single_click_execution}
\end{equation}
where $p_{\mathrm{det}}$ is the probability that an emitted photon is detected \cite{pompili2021realization}. When $p_{\mathrm{succ}} \ll 1$, as is typically the case for NV centres, it can be beneficial to perform attempts in batches \cite{pompili_experimental_2022} in order to avoid overhead in the software stack (the outcome of each attempt must be communicated to higher layers of the software stack). We now consider the case where one entanglement generation attempt corresponds to $M$ executions of the single-click protocol. If at least one of the $M$ execution succeeds, then the entanglement generation attempt is successful. The success probability of an individual attempt is therefore given by
\begin{equation}
    p = 1 - (1-p_{\mathrm{succ}})^M.
\end{equation}
Letting $\lambda \coloneqq 1/(2p_{\mathrm{det}} M)$, from (\ref{eqn:fidelity_single_click_execution}) we have that 
\begin{equation}\label{eq:full-trade-off}
    p = 1 - \left(1-\frac{ (1-F)}{\lambda M}\right)^M.
\end{equation}
Then, using the fact that $\lim_{K\rightarrow\infty}(1+\frac{x}{K})^K=e^x$, when $M\sim p_{\mathrm{det}}^{-1}$ is large the above is approximated as 
\begin{equation}
    p \approx 1 - e^{- \frac{1-F}{\lambda}},
\end{equation}
which results in the trade-off relation 
\begin{equation}
    F \approx 1 + \lambda \ln(1-p).
    \label{eqn:single_click_batched_tradeoff_appendix}
\end{equation}
The relation (\ref{eqn:single_click_batched_tradeoff_appendix}) is used in the illustration of our results in Section \ref{sec:performance_comparison}. The trade-off relation $F\approx 1-\lambda p$ that was found in \cite{davies_tools_2024} can be seen as the first-order approximation of (\ref{eqn:single_click_batched_tradeoff_appendix}) at $p=0$. 

We now show that (\ref{eqn:single_click_batched_tradeoff_appendix}) is an accurate approximation in the regime where $M\sim p_{\mathrm{det}}^{-1}$ is large.
To bound the error of the approximation, we use the following inequality:
\begin{equation}
    \left(1-\frac{1}{y} \right)^{y-1} > \frac{1}{e} > \left(1-\frac{1}{y} \right)^{y}, \;\;\; \text{ for } y>1.
    \label{eqn:inequality_for_tradeoff}
\end{equation}
The proof of (\ref{eqn:inequality_for_tradeoff}) is given in Appendix \ref{app:inequality_proof}. We recall that the true trade-off is given by 
\begin{equation}
    \tilde{p} = 1 - \left(1-\frac{ (1-F)}{\lambda M}\right)^M
\end{equation}
and the approximate trade-off 
\begin{equation}
    p = 1 - e^{- \frac{1-F}{\lambda}}.
\end{equation}
We firstly claim that $p<\tilde{p}$.  Letting $ r = (1-F)/\lambda$ and $y=M/r$, we have $r > 0$. Then, 
\begin{align}
    \tilde{p} - p &= e^{-r}  - \left(1-\frac{ r}{M}\right)^M  \\ &=  e^{-r} - \left(1-\frac{ 1}{y}\right)^{yr}  > 0, 
\end{align}
where in the final step we have made use of the upper bound from (\ref{eqn:inequality_for_tradeoff}). We now bound the difference $|p - \tilde{p}|$. We have 
\begin{align}
    |\tilde{p} - p| &= e^{-r} - \left(1-\frac{ 1}{y}\right)^{yr}  \\ &< e^{-r} - \left(e^{-1}\left(1-\frac{1}{y} \right) \right)^r \\ &= e^{-r}\left(1 - \left(1-\frac{1}{y} \right)^r \right) \\ &< 1 - \left(1-\frac{1}{y} \right)^r,
\end{align}
where in the second inequality we have used the lower bound from (\ref{eqn:inequality_for_tradeoff}), and in the final inequality we have used $e^{-r}<1$ (since $r>0$). In particular, making use of the power series expansion, we see that 
\begin{equation}
    |\tilde{p} - p| \approx \frac{r}{y} = \mathcal{O}\left( \frac{1}{M}\right) ,
\end{equation}
i.e. as long as $M \sim p_\mathrm{det}^{-1}$ is large, our approximation remains accurate.

\section{Proof of (\ref{eqn:inequality_for_tradeoff})}
\label{app:inequality_proof}
To prove the left inequality of (\ref{eqn:inequality_for_tradeoff}), we firstly take the logarithm of both sides, to obtain 
\begin{equation}
    (y-1)\ln \left(1-\frac{1}{y}\right)>-1, \text{   for }y>1.
    \label{eqn:left_hand_inequality}
\end{equation}
For ease of notation, we let 
\begin{equation}
    f(y) \coloneqq (y-1)\ln \left(1-\frac{1}{y}\right).
\end{equation}
To prove (\ref{eqn:left_hand_inequality}), we firstly claim that  
\begin{equation}
    \lim_{y\rightarrow \infty } f(y) = -1.
    \label{eqn:limit_y}
\end{equation}
Letting $z \coloneqq 1/y$, (\ref{eqn:limit_y}) is equivalent to 
\begin{equation}
    \lim_{z \rightarrow 0} \left(\frac{1-z}{z}\right)\ln\left(1-z\right) = -1,
\end{equation}
which one may verify with L'Hôpital's rule: differentiating both the numerator and denominator,
\begin{align}
    \lim_{z \rightarrow 0} \frac{(1-z)\ln\left(1-z\right)}{z}  &= \lim_{z \rightarrow 0} \frac{-\ln\left(1-z\right) -1 }{1} \\ &= -1.
\end{align}
Having shown (\ref{eqn:limit_y}), to prove (\ref{eqn:left_hand_inequality}) it suffices to show that the function $f(y)$ is decreasing. We have
\begin{align}
    \dv{f}{y} &= \ln \left(1-\frac{1}{y}\right) + (y-1)\frac{\frac{1}{y^2}}{1-\frac{1}{y}} \\ &= \ln \left(1-\frac{1}{y}\right)  + \frac{1}{y}.
\end{align}
Since
\begin{equation}
    \ln(1+z)<z \text{   for } z>-1,
    \label{eqn:inequality_z}
\end{equation}
by letting $y=-1/z$ it follows from the above that $\dv{f}{y}<0$, which suffices to prove the left inequality of (\ref{eqn:inequality_for_tradeoff}). 

We now show the right inequality of (\ref{eqn:inequality_for_tradeoff}), which taking the logarithm of both sides is equivalent to
\begin{equation}
    -\frac{1}{y} > \ln \left( 1-\frac{1}{y}\right), \text{   for } y>1.
\end{equation}
Again, letting $y = -1/z$, by (\ref{eqn:inequality_z}) this holds.

\bibliographystyle{IEEEtran}
\bibliography{references}

\vfill

\end{document}